\documentclass[a4paper,fleqn,usenatbib]{mnras}
\usepackage{mathrsfs}
\usepackage{color}
\usepackage{mathptmx}
\usepackage[T1]{fontenc}
\usepackage{ae,aecompl}
\usepackage{mathrsfs}
\usepackage{svg}

\usepackage{graphicx}	
\usepackage{amsmath}	
\usepackage{amssymb}	
\usepackage{csquotes}   
\DeclareMathAlphabet{\pazocal}{OMS}{zplm}{m}{n}

\newcommand{\lya}{Ly$\rm{\alpha}$~} 
\newcommand{\lyb}{Ly$\rm{\beta}$~} 

\newcommand{\mpc}{$h^{-1}{\rm Mpc}\,$}

\def\to{$\,T_{\rm 0}\,$}
\def\uo{$\,u_{\rm 0}\,$}
\def\tre{$\,T_{\rm reion}\,$}
\def\lmb{$\,\lambda_{\rm mfp}\,$}

\newcommand{\avgMFP}{\langle \lambda^{912}_{\rm mfp}\rangle}
\newcommand{\mfp}{\lambda^{912}_{\rm mfp}}

\def\taueff{$\tau_{\rm{eff}}\,$}
\newcommand{\zre}{$z_{\rm{re}}\,$}
\newcommand{\comment}[1]{}

\newcommand{\GammaHI}{$\Gamma_\mathrm{HI}$}
\newcommand{\Msun}{$M\textsubscript{\(\odot\)}$}
\newcommand{\HI}{\hbox{H$\,\rm \scriptstyle I\ $}}

\newcommand{\HeI}{\hbox{He$\,\rm \scriptstyle I\ $}}

\newcommand{\HeII}{\hbox{He$\,\rm \scriptstyle II\ $}}

\hypersetup{draft}

\begin{document}

\title[Observing the tail of reionization]{Observing the tail of reionization: neutral islands in the $z=5.5$ Lyman-$\alpha$ Forest}

\author[F. Nasir et al.]{Fahad Nasir$^{1}$\thanks{E-mail: fahadn@ucr.edu} and Anson D'Aloisio$^{1}$,
\\$^{1}$Department of Physics and Astronomy, University of California, Riverside, CA 92521, USA}


\label{firstpage}
\pagerange{\pageref{firstpage}--\pageref{lastpage}}
\maketitle

\begin{abstract}
Previous studies have noted difficulties in modeling the highest opacities of the $z>5.5$ Ly$\alpha$ forest, epitomized by the extreme Ly$\alpha$ trough observed towards quasar ULAS J0148+0600.  One possibility is that the most opaque regions at these redshifts contain significant amounts of neutral hydrogen.  This explanation, which abandons the common assumption that reionization ended before $z=6$, also reconciles evidence from independent observations of a significantly neutral Universe at $z=7.5$.  Here we explore a model in which the neutral fraction is still $\approx 10\%$ at $z=5.5$.  We confirm that this model can account for the observed scatter in Ly$\alpha$ forest opacities, as well as the observed Ly$\beta$ transmission in the J0148 trough.  We contrast the model with a competing ``earlier" reionization scenario characterized by a short mean free path and large fluctuations in the post-reionization ionizing background.  We consider Ly$\alpha$ and Ly$\beta$ effective optical depths, their correlations, trough size distributions, dark pixel fractions, the IGM thermal history, and spatial distributions of Lyman-$\alpha$ emitters around forest sight lines.   We find that the models are broadly similar in almost all of these statistics, suggesting that it may be difficult to distinguish between them definitively.   We argue that improved constraints on the mean free path and the thermal history at $z>5$ could go a long way towards diagnosing the origin of the $z>5.5$ opacity fluctuations.
\end{abstract}

 

\section{Introduction} \label{sec:intro}

Constraints on the timing and duration of reionization provide insight into the first sources of ionizing photons in the Universe.  The modern view is that the bulk of reionization occurred between $z\approx 6 - 10$.  This is based largely on the Thomson scattering optical depth ($\tau_{\rm es}$) constraints from the Planck satellite, which place the mid-point of reionization at $z_{\rm re}\simeq7.7\pm0.7$ \citep{Planck_2018}. 
Further evidence comes from the rapidly declining number density of \lya emitters (LAEs) at $z>6$
\citep[e.g.][]{Schenker_2012ApJ,Schmidt_2016ApJ,2018ApJ...856....2M}, as well as damping wing features and small proximity zones in the spectra of the highest redshift quasars \citep{Mortlock_2011Nature,2018Natur.553..473B, Davies_2018ApJ}.  On the other hand, a common working assumption, based on the prevalence of Ly$\alpha$ forest transmission, is that reionization was complete before $z=6$ \citep{Songaila_2004AJ,Fan_2006AJ,McGreer_2015MNRAS}.  Here we will consider the possibility that reionization ended significantly later than $z=6$. 

More potential evidence for late reionization comes from large opacity fluctuations in the $z>5$ \lya forest.   A sharp increase in the overall opacity, and in its scatter on $50 h^{-1}$ Mpc scales, was observed in \citet{Fan_2006AJ}, and later confirmed by others \citep{Becker_2015MNRAS,Bosman_2018MNRAS,Eilers_2018ApJ}. While the rapid evolution in the overall opacity is not by itself an unambiguous signature of reionization \citep{2011ApJ...743...82M}, the scatter is harder to explain otherwise \citep[for some attempts, see][]{D'Aloisio_2015ApJ, Davies_2016MNRAS, Chardin_2017MNRAS}.    

Of particular importance is the $110h^{-1}$ Mpc long \lya trough centered at $z\approx 5.7$ in the sightline of quasar ULAS J0148+0600 (henceforth J0148; \citealt{Becker_2015MNRAS}).  Interestingly, the trough exhibits significant transmission in the coeval Ly$\beta$ forest, which seemed to reaffirm that reionization ended by $z = 6$. A key insight came when the test of \citet{2018ApJ...860..155D} was carried out by \citet{Becker_2018ApJ}.  They measured the spatial distribution of $z=5.7$ LAEs centered on the J0148 trough, probing the large-scale density of the region.  The observed deficit of LAEs within $20 h^{-1}$ Mpc of J0148 suggests that the sightline passes through a cosmic void.  This interpretation is supported by a deficit of Lyman Break Galaxies around the sight line \citep{2019arXiv190909077K}.    

The prevailing model after \citet{Becker_2018ApJ} was that of \citet{Davies_2016MNRAS}, in which in which the opacity variations are driven by large fluctuations in the post-reionization ultraviolet background (UVB).   The measurements of \citet{Becker_2018ApJ}, however, also raised interesting questions in the context of late reionization.  If reionization ended around $z=6$, then the J0148 void was likely reionized not much earlier than $z=6$. In this case, the void should have been hot ($T \sim 20,000$ K), which would make more extreme the low level of UVB needed to reproduce the Ly$\alpha$ trough, because the amount of absorbing \HI scales as $T^{-0.7}$ \citep{D'Aloisio_2015ApJ, D'Aloisio_2019ApJ}. Also, lowering the mean free path significantly, as is required in the model of \citealt{Davies_2016MNRAS}, increases proportionately the ionizing emissivity that is needed to complete reionization \citep{D'Aloisio_2018MNRAS}, perhaps requiring larger escape fractions.

An alternative model explored in \citet{Kulkarni_2019MNRAS} and \citet{Keating_2019arXiv} circumvents these issues by abandoning the assumption that reionization ended by $z=6$.  In their model, reionization ends at $z\approx 5.2$ such that the voids still contain significant amounts of neutral hydrogen at $z>5.5$. The ``neutral islands" are the essential new ingredient for reproducing the high opacity tail of the Ly$\alpha$ forest fluctuations.  Notably, this model is consistent with evidence of a significantly neutral IGM at $z=7.5$, and it naturally produces long \lya troughs like the one in J0148 \citep{Keating_2019arXiv}.  If confirmed, the model places reionization squarely in the regime where its tail end could potentially be studied directly with existing \lya and Ly$\beta$ forest spectra.  

In this paper we further study the scenario in which reionization was still ongoing at $z=5.5$.    In contrast to the radiative transfer simulations of \citet{Kulkarni_2019MNRAS} and \citet{Keating_2019arXiv}, we employ approximate ``semi-numeric" methods to model reionization's effects on the forest.  We will exploit the flexibility and inexpensiveness of these methods to explore the parameter space of reionization, adding to the previous studies.  A basic question that we aim to address here is as follows.  Consider two scenarios that both reproduce the large scatter in the $z>5$ forest.   In one scenario, the scatter is driven by neutral islands from ongoing reionization \citep{Kulkarni_2019MNRAS, Keating_2019arXiv}.  In the other, it is driven by large UVB fluctuations in the {\it post-reionization} IGM \citep{Davies_2016MNRAS, D'Aloisio_2018MNRAS}. Can we easily distinguish between these scenario with existing data?

The remainder of this paper is organized as follows. In \S \ref{sec:methodology}, we describe the simulations and modeling methods used in this work. In \S \ref{sec:models}, we present our reionization models and provide physical insight into their features, including the existence of long Ly$\alpha$ troughs. In \S \ref{sec:testing}, we explore statistical tests to distinguish between the models. 
Lastly, we summarize our results in \S \ref{sec:conclusions}. Throughout this paper we use comoving distances unless otherwise stated. We adopt cosmological parameters of $\Omega_{\rm m}=0.31$, $\Omega_{\rm \Lambda}=0.692$, $h=0.68$, $\Omega_{\rm b}=0.048$, $\sigma_{\rm 8}=0.82$ and $n=0.97$  consistent with the Planck measurements \citep{Planck_2018}.


\begin{figure}
\begin{center}
\resizebox{8.5cm}{!}{\includegraphics{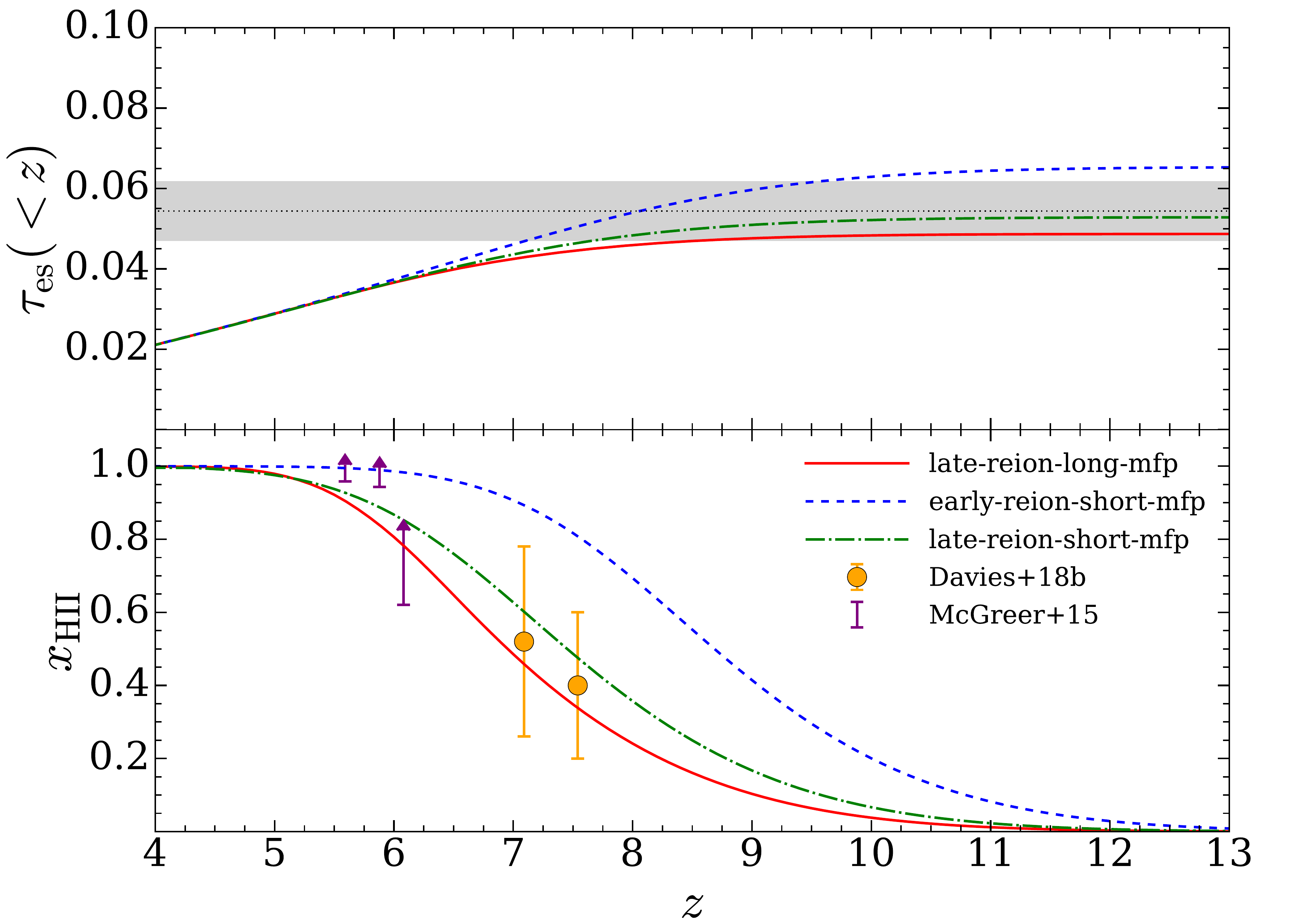}}
\end{center}
\caption{Reionization histories of the models used in this work. {\it Top:} the integrated electron scattering optical depth parameter, $\tau_{\rm es}$. The grey shading spans the $1\sigma$ limits from \citet{Planck_2018}. {\it Bottom:} The redshift evolution of the volume-weighted average \HI\ fraction. For reference, we show constraints from the dark pixel analysis of \citet{McGreer_2015MNRAS} and the damping wing analysis of \citet{Davies_2018ApJ} using high-$z$ quasars at $z\simeq7.1$ and $7.5$.}
\label{FIG:xHII_taues}
\end{figure}

\begin{figure*}
\resizebox{17cm}{!}{\includegraphics{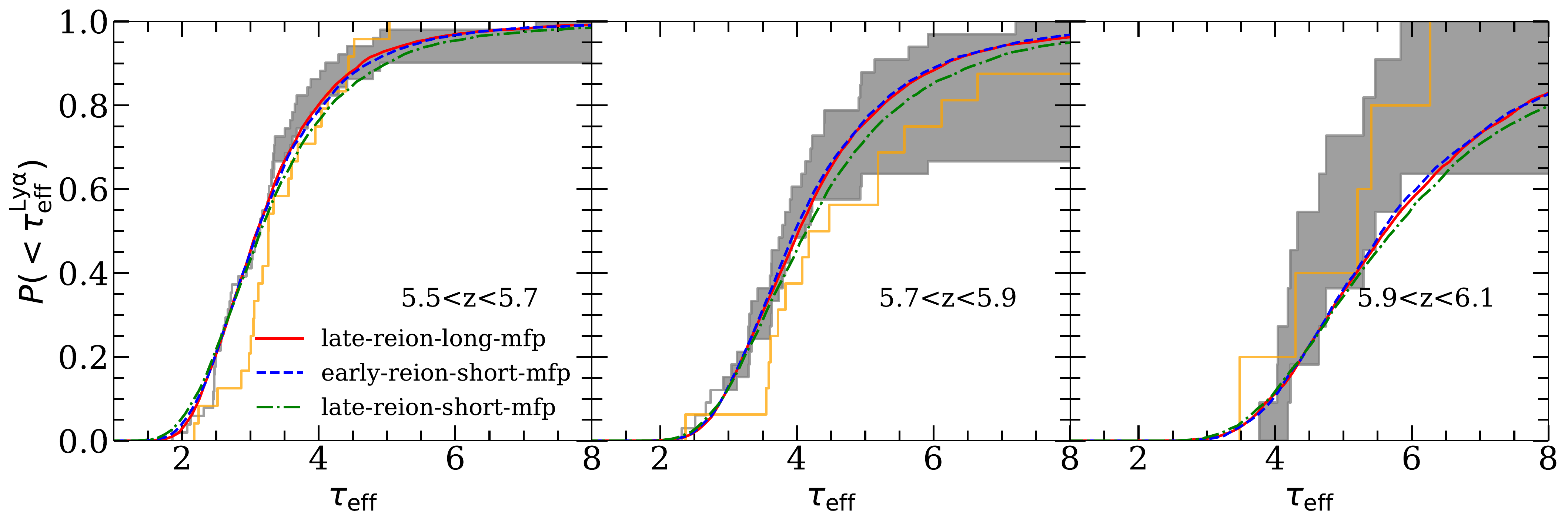}}
\caption{Comparison of the Ly$\alpha$ forest effective optical depth (\taueff) distributions in our models.  The grey histograms show the measurements of \citet{Bosman_2018MNRAS}.  The shading spans their ``optimistic" and ``realistic" treatments of non-detections.  The orange histograms show the measurements of \citet{Eilers_2018ApJ}.  Note that we have calibrated our models to match the mean fluxes reported by \citet{Bosman_2018MNRAS}.}
\label{fig:taueff_lya}
\end{figure*}

\section{Numerical Methodology}
\label{sec:methodology}

We post-process a cosmological hydrodynamics simulation to model large-scale fluctuations in the \lya forest.  Here we describe the elements of our numerical modeling.  Readers uninterested in these technical details may skip to \S \ref{sec:models} without loss of continuity.

\subsection{Hydrodynamics simulation}

The basis of our models is an Eulerian hydrodynamics simulation that we ran with a modified version of the code of \citet{Trac_2004NewA}.   The run has $2\times2048^3$ gas and dark matter resolution elements in a box with $L =200$ \mpc.   The simulation was initialized at $z=300$ using first-order Lagrangian perturbation theory and transfer functions generated by \textsc{CAMB} \citep{Lewis_2000ApJ}.  The gas was nominally flash ionized to a temperature of 20,000 K at $z=7.5$ and a uniform ionizing background with $J_{\rm \nu} \propto \nu^{-1.5}$ between 1 and 4 Ry was applied thereafter (here $\nu$ is frequency).  The strength of the background is tuned such that the hydrogen photoionization rate is $\Gamma_{\mathrm{HI}}=10^{-13} \mathrm{s}^{-1}$, but in the ensuing sections we will describe how we post-process the simulation to model fluctuations in $\Gamma_{\mathrm{HI}}$, as well as in the gas temperature.  The hydro simulation setup is nearly identical to that in \citet{D'Aloisio_2018MNRAS}.

\subsection{Semi-numerical reionization simulation}

We apply the excursion set model of reionization (\textsc{ESMR}) \citep{Furlanetto_2004ApJ,Mesinger_2011MNRAS} to obtain realizations of the reionization redshift field, \zre$(\mathbf{x})$. We filter the initial density field\footnote{We have also tested applying the ESMR to the evolved density field at $z=5.8$.  We found that our results/conclusions are insensitive to the effects of density evolution on our ESMR models. } of our hydro simulation on a hierarchy of scales and compute the local collapsed fraction using the conditional Press-Schechter expression, $f_{\rm coll} = \mathrm{erfc} \left[ (\delta_c(z) - \delta_m)/\sqrt{2 (\sigma^2(m_\mathrm{min}) - \sigma^2(m))}\right]$, where $\delta_c(z)$ is the critical density in the spherical collapse model, $\delta_m$ is the mean linear density contrast within a spherical top-hat containing mass $m$, and $\sigma^2(m)$ is the variance of linear density fluctuations on scale $m$ \citep{Schechter_1974ApJ,1991ApJ...379..440B,1994MNRAS.271..676L}.    We flag a cell as ionized once the condition $\zeta f_{\rm coll} \geq 1$ is met, where $\zeta$ is a constant ionizing efficiency parameter that encapsulates the efficiency at which collapsed baryons form stars and deliver ionizing photons to the IGM.  The minimum halo mass for forming galaxies is assumed to be $m_\mathrm{min} = 2\times10^9$ \Msun, and we tune $\zeta$ to achieve the desired reionization histories (see \S \ref{sec:models}).    The mean free path is modeled in a simplistic way by adopting the maximum filter scale as a free parameter \citep{Furlanetto_2005MNRAS,Mesinger_2011MNRAS}.  We find that imposing a mean free path in the ESMR, albeit crudely, is required to produce reasonable ionization front (I-front) speeds ($\lesssim 10^4$ km/s) toward the end of reionization \citep[see][]{D'Aloisio_2019ApJ,2019A&A...622A.142D}.  Our reionization models are generated on a coarser grid of size $N=256^3$, with a spatial resolution of $\Delta x = 0.78$ \mpc.  This is sufficient for our purposes given that the approach used here is unlikely to capture the detailed structure of reionization on scales $\lesssim 1$ \mpc.   

\subsection{Temperature fluctuations from reionization}

To model the imprint of reionization on IGM temperatures, we adopt an approach similar to that of \citet{D'Aloisio_2019ApJ} (see also \citealt{Sanderbeck_2016MNRAS, 2018ApJ...860..155D}).  The temperature evolution of each gas parcel with over-density, $\Delta$, after it has been reionized, is governed by\footnote{Note that this equation assumes that the total number of particles remain fixed, a good approximation after reionization.} \citep[e.g.][]{Miralda_1994MNRAS,McQuinnSanderbeck2016}

\begin{equation}
 \frac{dT}{dt} = \frac{2\mu m_H}{3k_B\rho}(\mathscr{H}-\Lambda)+\frac{2T}{3\Delta}\frac{d\Delta}{dt}-2HT,
 \label{eq:temp}
\end{equation}

\noindent
where $H$ is the Hubble parameter, $\mathscr{H}$ is the sum of the photo-heating rates, and $\Lambda$ contains the radiative cooling rates.  The second term on the right describes adiabatic heating and cooling from structure formation. The final term arises from adiabatic cooling due to the expansion of the Universe. 
Strictly speaking, equation (\ref{eq:temp}) applies in the Lagrangian picture but we use it here to describe the temperature evolution of our Eulerian gas cells.  We aim to model the large-scale structure of the temperature field during/after reionization, and not the details on small scales. \citet{D'Aloisio_2019ApJ} found that the above approximation works well for this purpose when compared against a fully coupled radiative hydrodynamics simulation. 

For each of our reionization models, we use the \zre\ field to compute the I-front velocity field, $v_{\rm IF}$, for each cell using the gradient method of \citet{D'Aloisio_2019ApJ} (see also \citealt{2019A&A...622A.142D}).  Each cell is then initialized to a post-I-front temperature, \tre$(v_{\rm IF})$, at the appropriate \zre\ using the fitting formula of \citet{D'Aloisio_2019ApJ}.  The subsequent evolution is obtained by solving the Eq.~(\ref{eq:temp}) numerically assuming a post-reionization ionizing background with $J_\nu \propto \nu^{-1.5}$ between 1 and 4 Ry.\footnote{Note that the post-reionization temperature evolution is not sensitive to the normalization of the ionizing background, as the photoheating and cooling rates (for highly ionized gas) do not depend on the normalization.}  \citet{McQuinnSanderbeck2016} showed that the temperature evolution in the diffuse IGM is relatively insensitive to the evolution of $\Delta$ under realistic assumptions.  Here we adopt the Zel'dovich pancake approximation to propagate backwards or forward in time the $z=5.8$ density values from our hydro simulation.   We adopt a fixed temperature of $T=10K$ for pre-reionization cells, but all of our results pertaining to observable quantities are insensitive to this choice.  Our temperature calculations are performed at the resolution of our hydro simulation, but note that the \zre fields, and therefore the \tre fields, are at the coarser resolution of $N=256^3$.

\subsection{UVB fluctuations}

Our models also include UVB fluctuations with a version of the \citet{Davies_2016MNRAS} model, modified to achieve higher resolutions and to approximate the radiative transfer shadowing effects around neutral islands.  For late reionization models, we will find that the shadowing is crucial for generating large opacity fluctuations.  The basic setup is similar to that in \citet{D'Aloisio_2018MNRAS}.   Halos were identified using a spherical over-density criterion of 200 times the cosmic mean matter density. We then abundance matched the halos to an interpolation of the rest-frame UV luminosity functions of \citet{Bouwens_2015ApJ}.  We assume that all halos with masses above $2\times 10^{10}~h^{-1} M_\odot$ (with at least $\approx 300$ dark matter particles) host a star forming galaxy.  This limit was chosen for completeness of the halo mass function, i.e. faithfully reproducing the mass function measured from the SCORCH I suite of N-body simulations, which includes runs that extend well beyond our resolution limit \citep{2015ApJ...813...54T}.  Note that, owing to the resolution of our hydro simulation, the minimum mass for these calculations is necessarily different than the $M_{\rm min} = 2\times 10^9$ $M_\odot$ used for the ESMR calculations.  We do not expect this to change our main conclusions, however, as most of the effects of larger $M_{\rm min}$ could be mimicked by using a somewhat lower mean free path.

To model the spatially varying mean free path, we bin the sources onto a coarse grid with $N=64^3$ and iteratively solve for $\Gamma_{\rm HI}(\mathbf{x})$ and \lmb$(\mathbf{x})$ under the assumption that the latter scales as \lmb$\propto {\Gamma}_{\rm HI}^{2/3} / \Delta$.  This scaling is motivated by analytic models \citep{Miralda-Escude_2000ApJ,Furlanetto_2005MNRAS} and by the scaling measured in radiative transfer simulations \citep{2011ApJ...743...82M}.\footnote{Strictly speaking, the scaling measured by \citet{2011ApJ...743...82M} applies in the limit of fixed distribution function in $\Delta$. It is therefore unclear whether our model for the scaling is appropriate for epochs near the end of reionization, when different regions were at different states of dynamical relaxation (as a result of the patchy heating during reionization).  However, for our fiducial late reionization model we will see that it is the neutral islands as well as their shadowing of the UVB that drives the forest opacity fluctuations, and not the variations from the $\lambda_{\rm{mfp}} \propto {\Gamma}_{\rm HI}^{2/3}$ scaling, suggesting that our main conclusions are insensitive this assumption.}  Our main upgrade to the approach of \citet{Davies_2016MNRAS} is that we add a second step in which we use the coarse $N=64^3$ \lmb\ field together with a direct sum over sources to recalculate $\Gamma_{\rm HI}(\mathbf{x})$ on a higher resolution grid with $N=256^3$ (the same resolution as our ESMR grids).  In this step, we remove ionizing photon contributions from sources whose sight lines pass through neutral islands.  We will see that this shadowing by the neutral islands leads to suppressed \GammaHI\ in their vicinity.  In what follows, we will characterize our models with $\avgMFP(z)$,  the volume-weighted average over our mean free path fields. (Here the ``912" denotes 912 \AA.) In all of our models, we set the maximum filter scale in our ESMR calculations to $\avgMFP(z=5.6)$ in an attempt to make the reionization models more consistent with the above fluctuating UVB models.   Lastly, we note that our approach does not capture the effects of the shadowing on the local mean free path near the neutral islands.  In reality, the \GammaHI\ around the islands may be even lower from this effect.

\subsection{Synthetic Ly$\alpha$/$\beta$ forest sight lines}
\label{sec:methods_skewers}

Finally, to construct synthetic quasar absorption spectra, we trace $4000$ sight-lines at random angles through the simulation box, each with length of 500\mpc (making use of the periodic boundary conditions). The neutral hydrogen densities for the ionized gas along the skewers are rescaled using our fluctuating temperature and $\Gamma_{\rm HI}$ fields under the assumption of photoionization equilibrium, and we set cells to be fully neutral if $z > z_{\rm re}$ according to the ESMR grids.  We calculate Ly$\alpha$ and Ly$\beta$ opacities along the sight lines using the (accurate) approximation of \citet{2006MNRAS.369.2025T} to the Voigt profile.  We iteratively adjust the normalization of \GammaHI$(\mathbf{x})$ in the \emph{ionized} regions to match the mean Ly$\alpha$ forest flux with observed values.  Unless otherwise noted, we match to the mean fluxes reported by \citet{Bosman_2018MNRAS}.  For some calculations (e.g. long Ly$\alpha$ troughs), we account for evolution in \GammaHI\ along the skewers by fitting a 4th order polynomial to the normalization of \GammaHI\ between $z=5.5-6$ and rescaling the skewers as a function of redshift. We model instrumental noise by drawing pixel-by-pixel from a Gaussian distribution according to the desired level of signal to noise ($S/N$).  Because this parameter depends on the particular comparison with observations, we provide these details below.

\begin{table*}
\caption{ Summary of models used in this work. The mean free paths are quoted in units of comoving $h^{-1}$Mpc (see text for a definition of these average quantities).   The middle three columns give volume-weighted mean \HI fractions at select redshifts.  The right-most columns give the mean \GammaHI\ in units of $\times 10^{-12}$ s$^{-1}$. }
\centering
 \begin{tabular}{lccccccccc}
  \hline
  \hline
  Model &         & $\avgMFP$ &         &         & $\langle x_{\rm HI} \rangle$ &  & & $\langle \Gamma_{-12} \rangle$ &      \\  \hline
        & $z=5.6$ & $z=5.8$   & $z=6.0$ & $z=5.6$ & $z=5.8$      & $z=6.0$ & ${z=5.6}$ & ${z=5.8}$  & ${z=6.0}$ \\ \hline
  {\UrlFont late-reion-long-mfp}.  & 30 & 27 & 23 &  0.097 & 0.141 & 0.194 & { 0.39} & { 0.25} & { 0.15}    \\
  {\UrlFont early-reion-short-mfp} & 10 & 9 & 8 &  0.005 & 0.009 & 0.014    & { 0.54}  & { 0.33} &  { 0.19}   \\
  {\UrlFont late-reion-short-mfp}  & 10 & 9 & 8 &  0.074 & 0.100 & 0.133     & {  0.48} & { 0.29} &  {  0.17}     \\\hline 
\end{tabular}
\label{tab:models}
\end{table*}

\section{Models of Reionization}

\label{sec:models}

\begin{figure*}
\begin{center}
\resizebox{5.8cm}{!}{\includegraphics{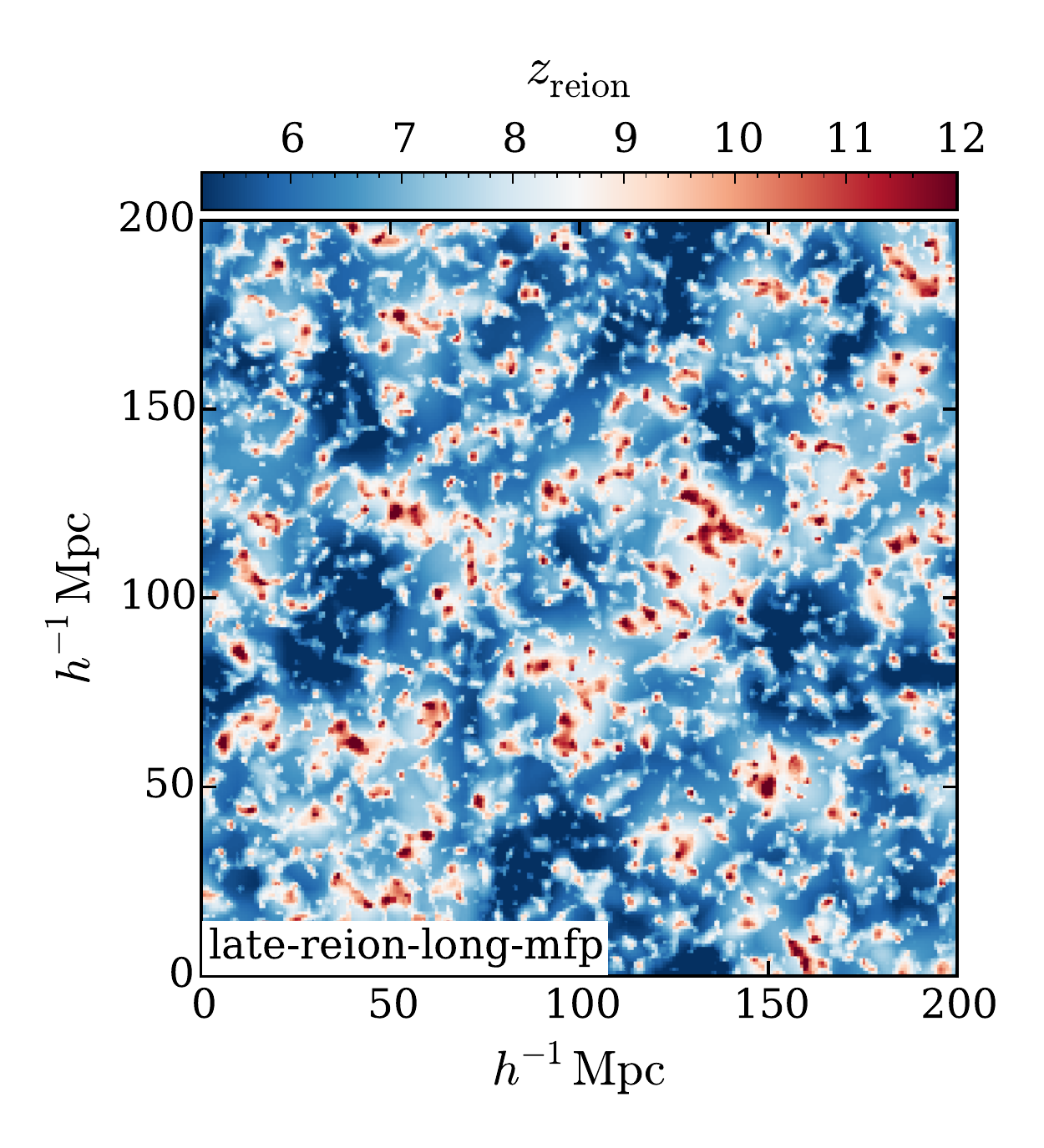}}
\hspace{-0.28cm}
\resizebox{5.8cm}{!}{\includegraphics{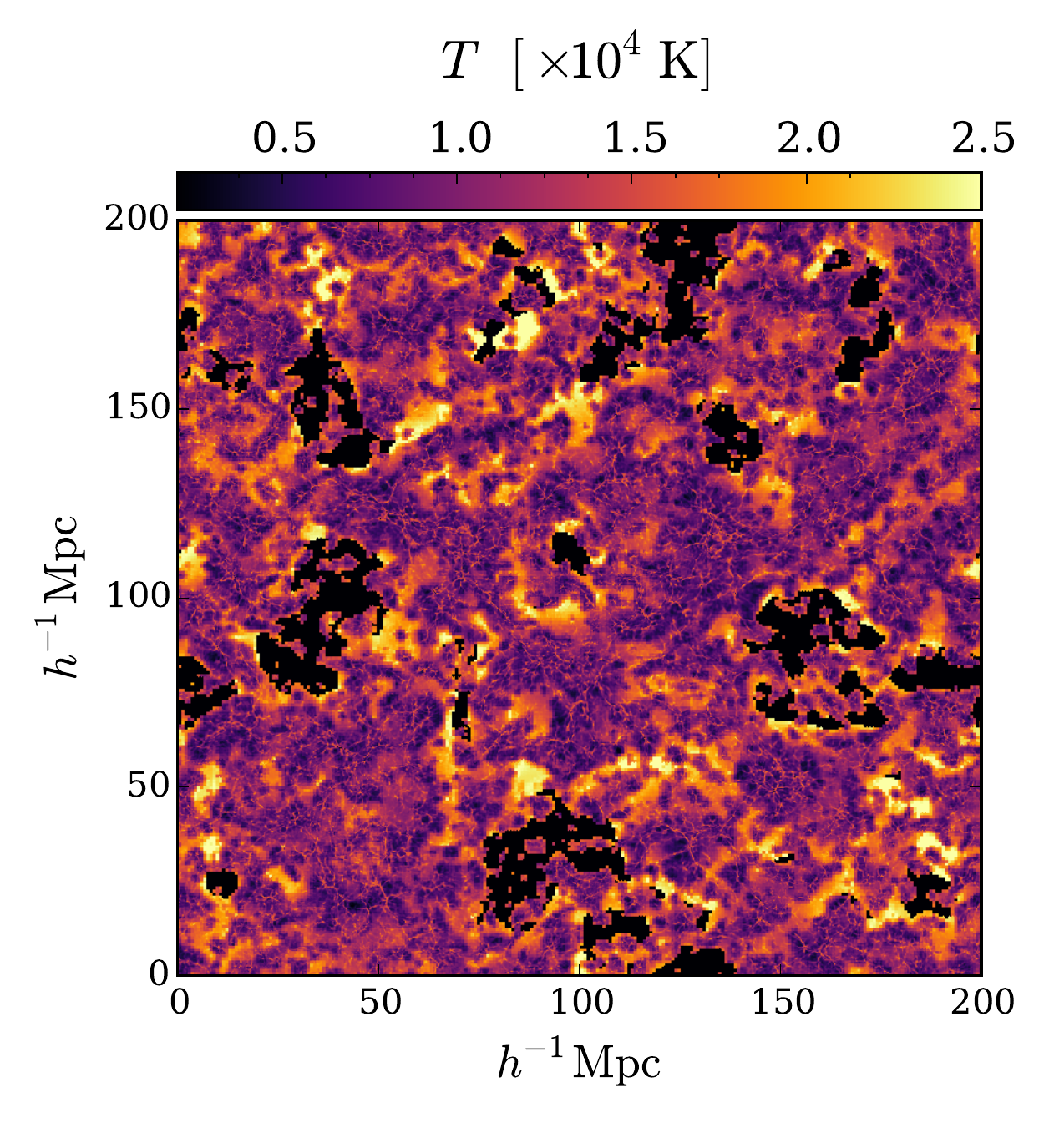}}
\hspace{-0.28cm}
\resizebox{5.8cm}{!}{\includegraphics{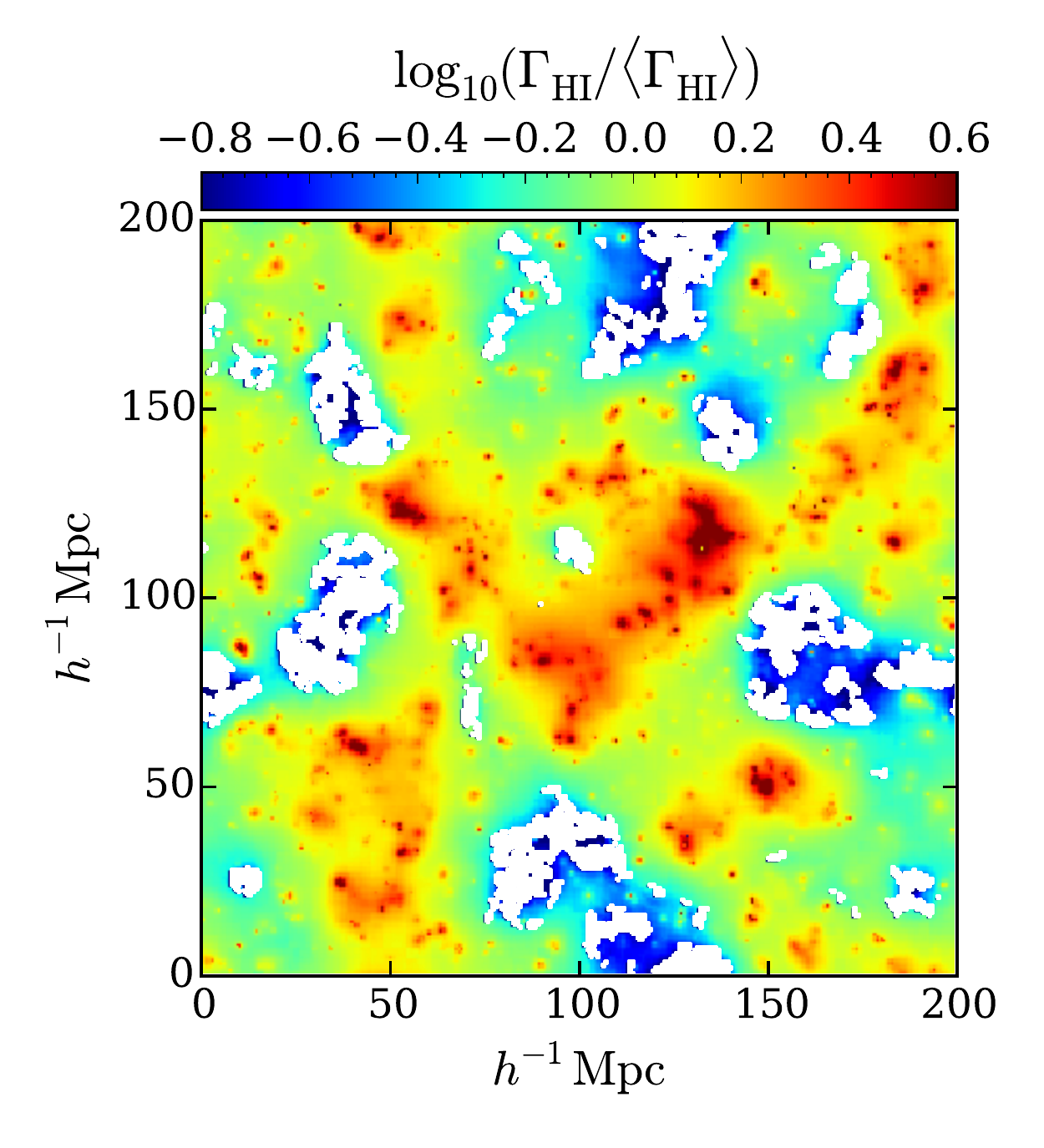}}
\hspace{-0.28cm}
\resizebox{5.8cm}{!}{\includegraphics{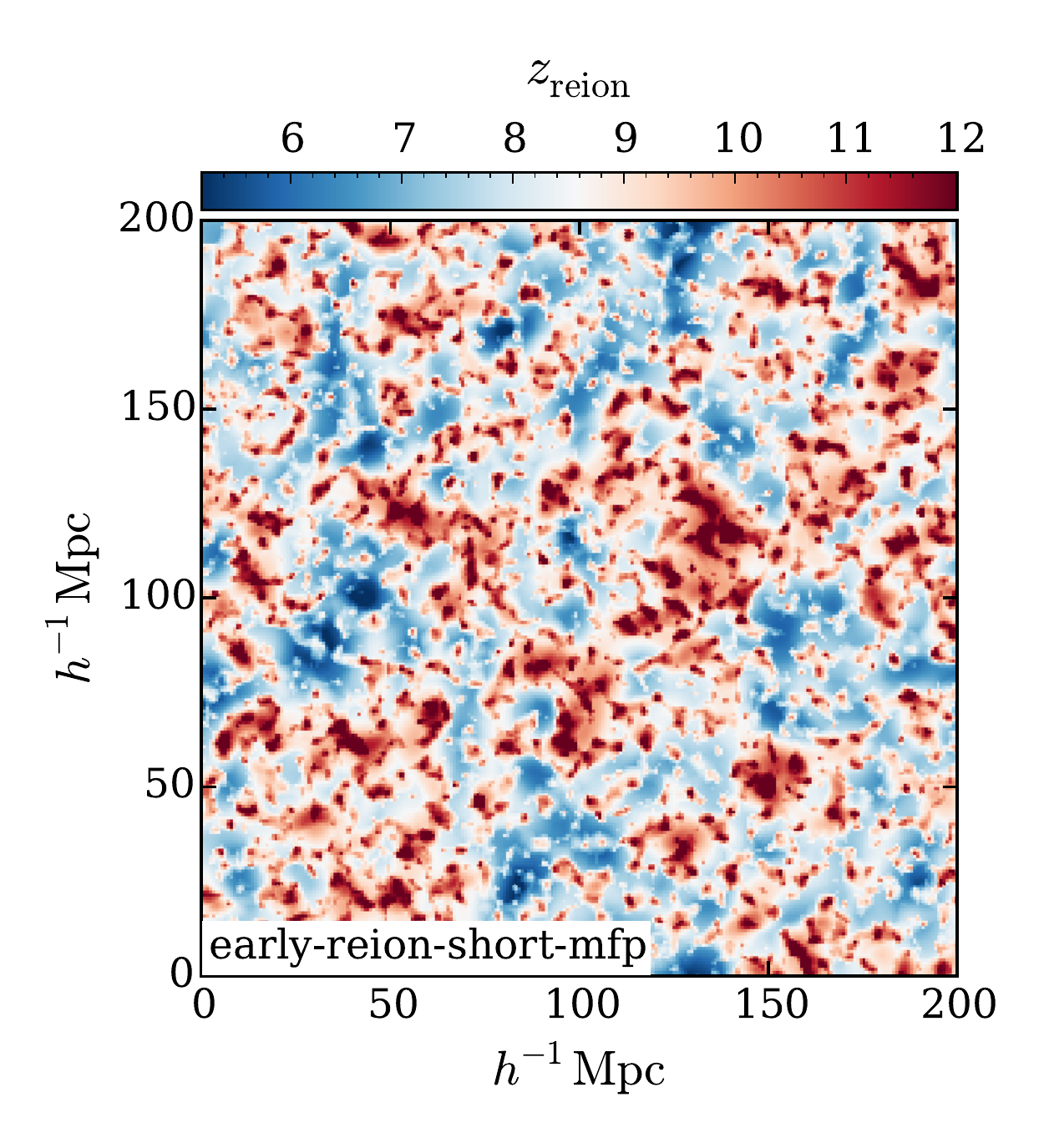}}
\hspace{-0.28cm}
\resizebox{5.8cm}{!}{\includegraphics{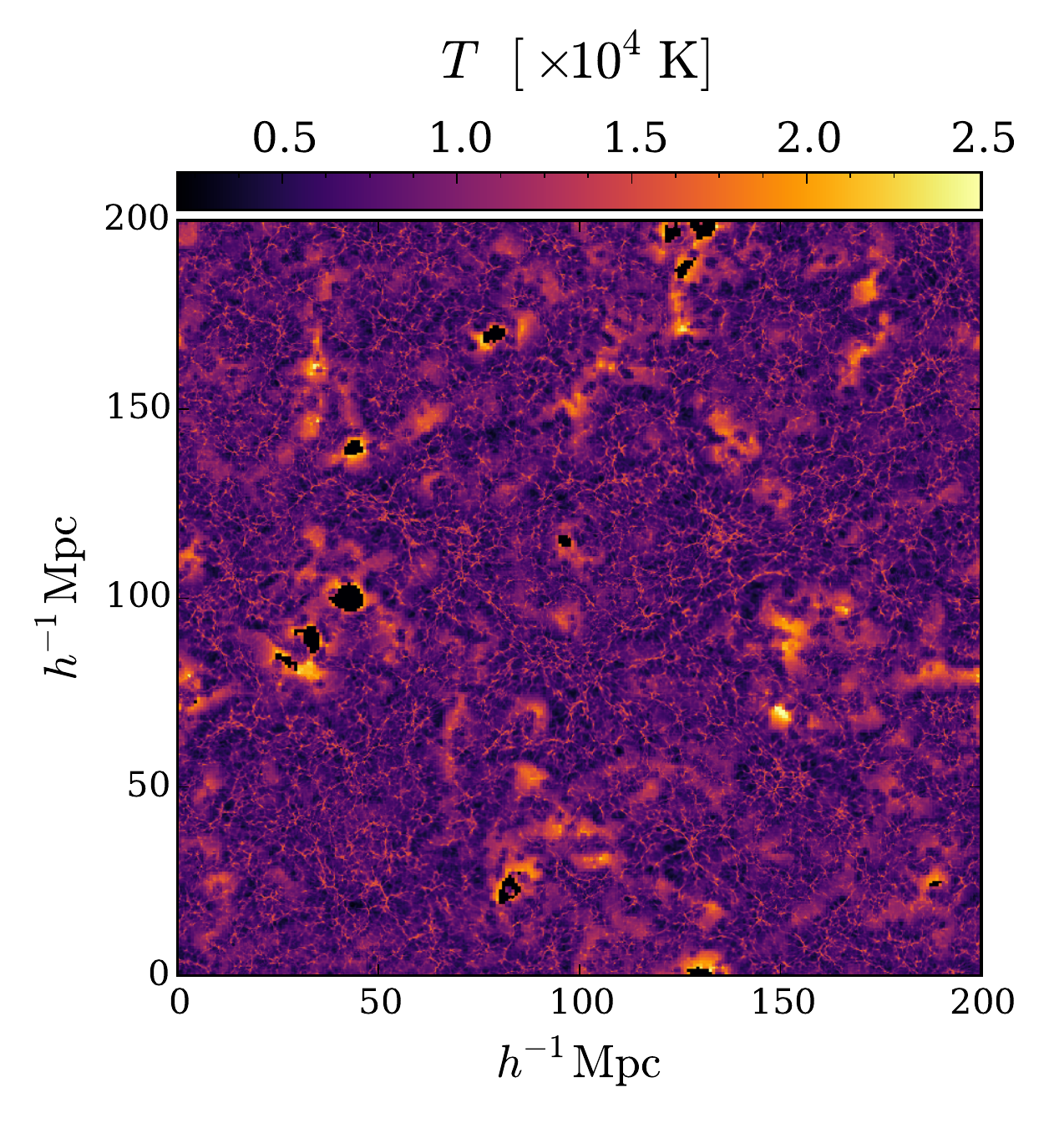}} 
\hspace{-0.28cm}
\resizebox{5.8cm}{!}{\includegraphics{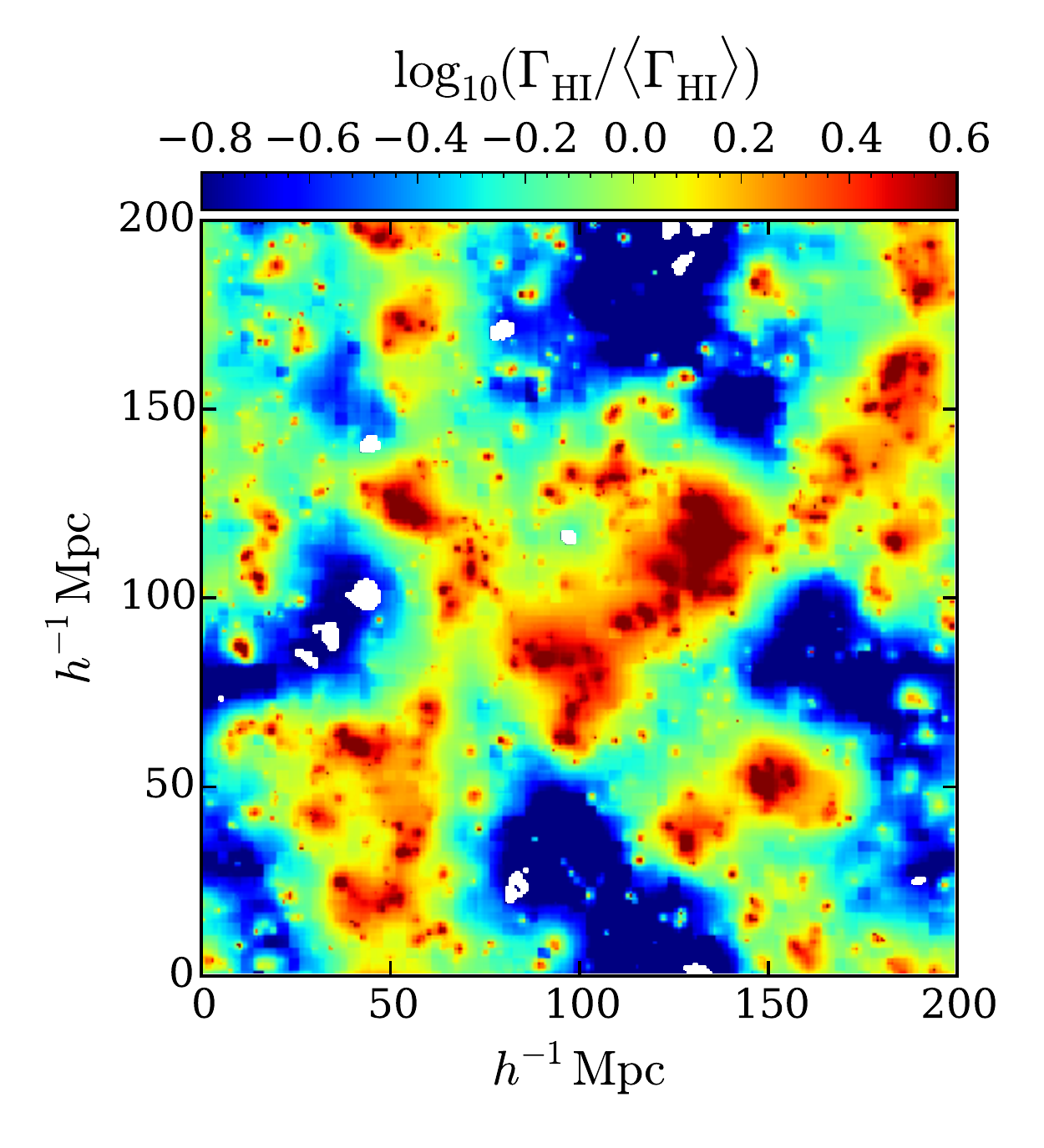}}
\hspace{-0.28cm}
\resizebox{5.8cm}{!}{\includegraphics{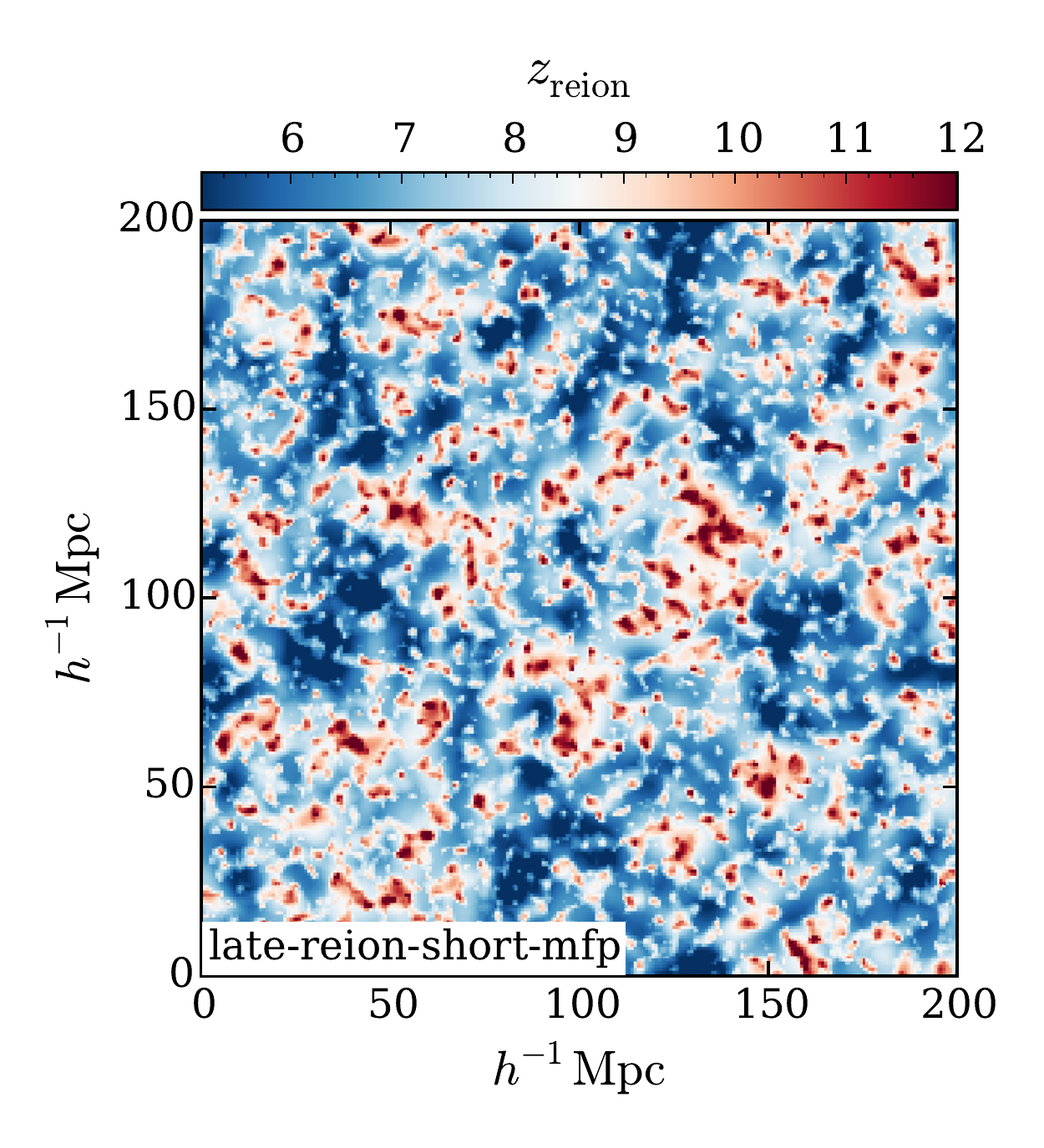}}
\hspace{-0.28cm}
\resizebox{5.8cm}{!}{\includegraphics{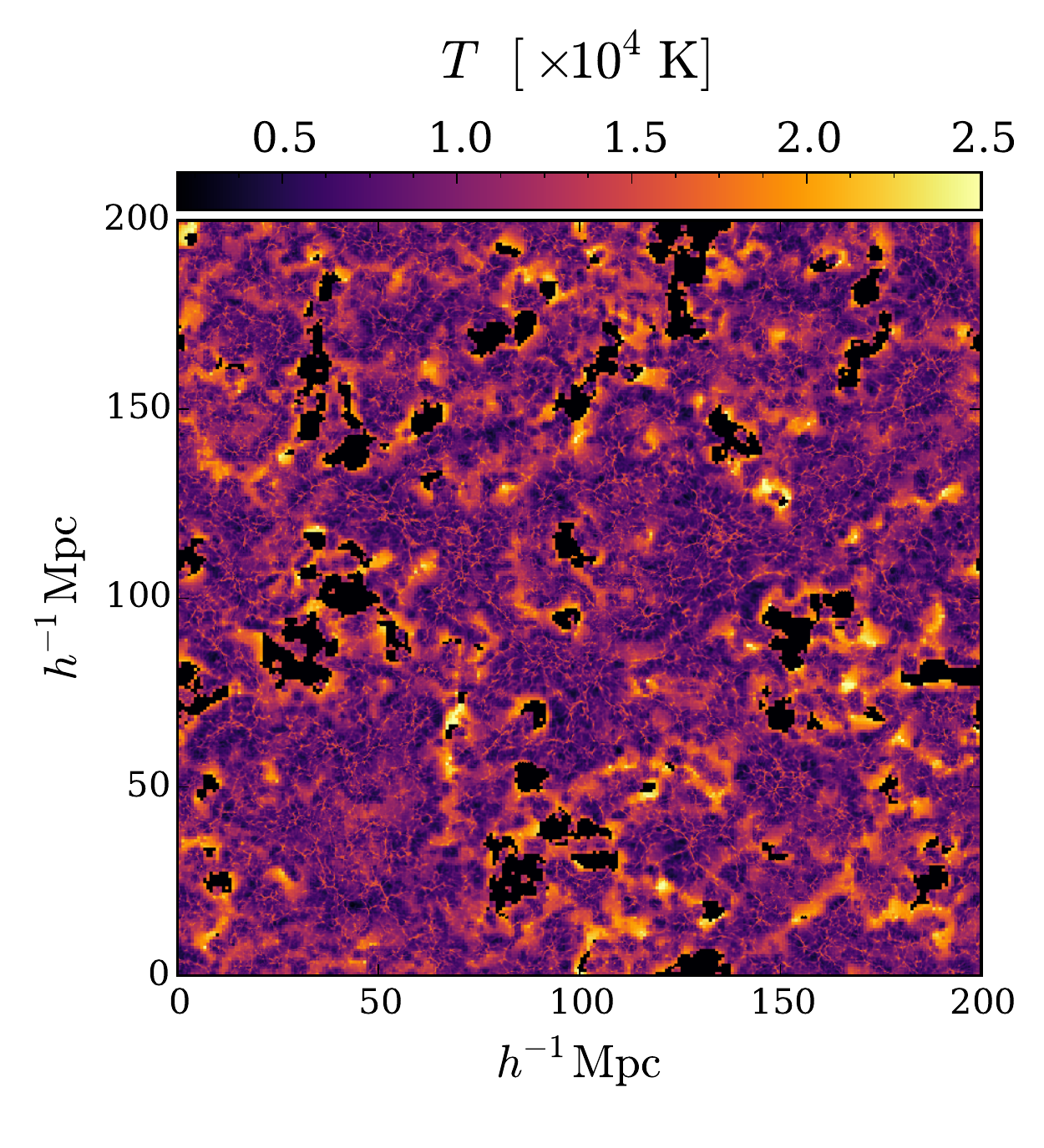}} 
\hspace{-0.28cm}
\resizebox{5.8cm}{!}{\includegraphics{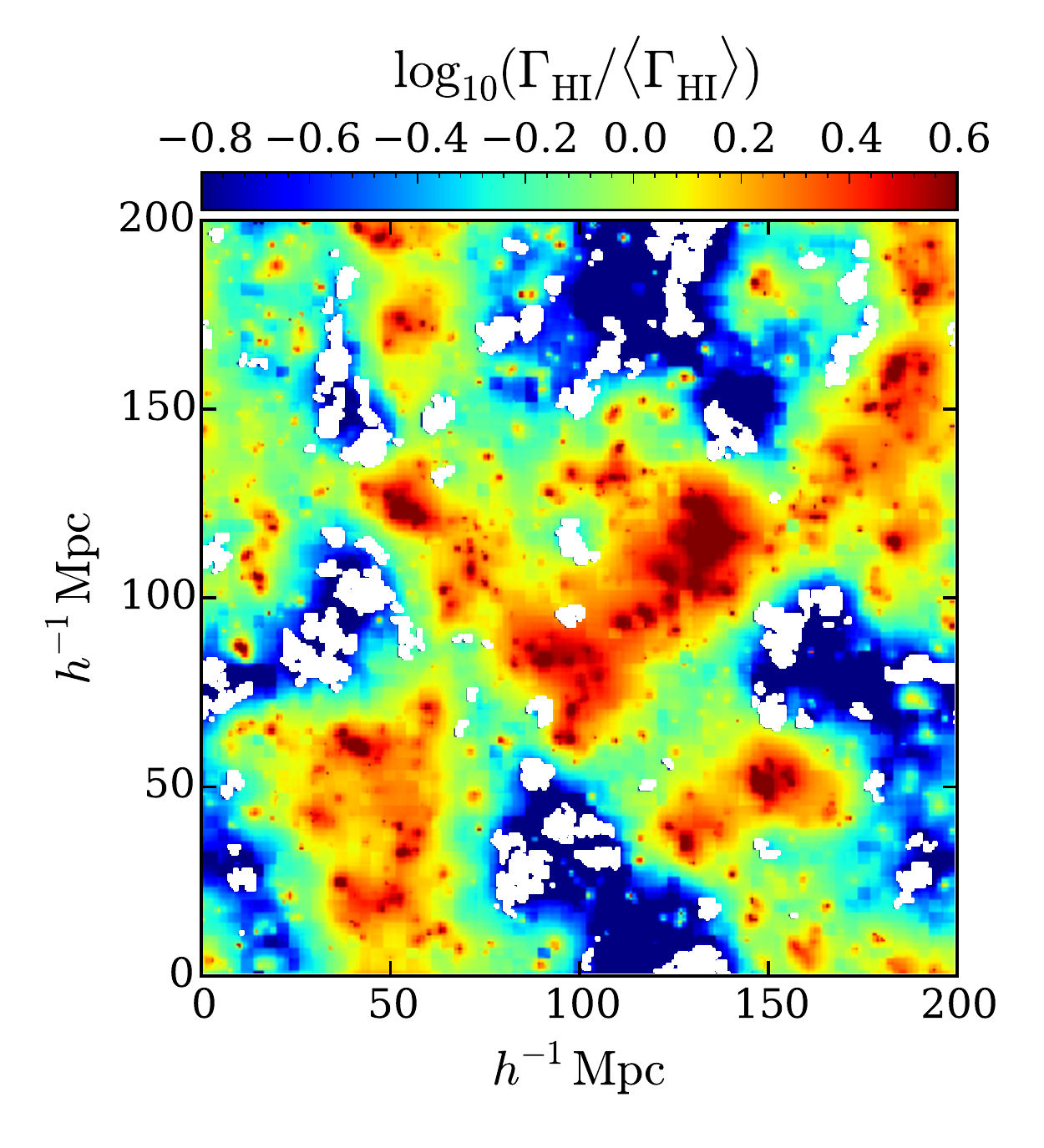}}
\hspace{-0.28cm}
\end{center}
\caption{Slices through our simulations at $z=5.6$, showing the interplay between fluctuations in our models. From top row to bottom: {\UrlFont late-reion-long-mfp}, {\UrlFont early-reion-short-mfp} and {\UrlFont late-reion-short-mfp}. The left, middle, and right columns correspond to slices through the reionization redshift field, the temperature field, and the \HI\ photoionization rate.  Neutral islands are shown in black (white) in the middle (right) panels.}
\label{fig:vis}
\end{figure*}

\subsection{Reionization scenarios}
\label{sec:scenarios}

We model three broadly different reionization scenarios to explore whether they might be immediately distinguishable with current observations.  Our models are characterized by the timing of reionization's end and by $\avgMFP(z)$.  The mean free path at $z>5$ is highly uncertain and even today's large-volume radiative transfer simulations of reionization likely lack the resolution to model it accurately.  Here we vary $\avgMFP(z)$ by a factor of 3 amongst our models to explore the widely open parameter space.     

The reionization histories of our models are shown in the bottom panel Fig. \ref{FIG:xHII_taues}.  For reference, we also show the $1\sigma$ constraints of \citet{McGreer_2015MNRAS} and \citet{Davies_2018ApJ}.  The top panel shows the corresponding cumulative electron scattering optical depths, $\tau_{\rm es}(<z)$, compared against the 1$\sigma$ limits measured by Planck \citep{Planck_2018}. Table \ref{tab:models} quantifies the mean free path and timing of reionization in our models.  For simplicity, we adopt a redshift dependence of $\mfp \propto (1+z)^{-4.4}$, in accordance with the fitting formula of \citet{Worseck_2014MNRAS}, which captures the trend of measurements at $z\lesssim 5.2$.  For reference, extrapolating the fit of \citet{Worseck_2014MNRAS} to higher redshifts yields $\mfp=38, 34$ and $30$ \mpc\  at $z=5.6, 5.8$ and $6.0$, respectively.   

Our fiducial model (red/solid) is a late reionization scenario with a significant amount of neutral hydrogen remaining in the IGM at $z=5.5$, and with a (relatively) longer $\avgMFP$.  We will see that large opacity fluctuations are driven by the presence of neutral islands in this model, which we denote as {\UrlFont late-reion-long-mfp}.  We will contrast this scenario against an updated version of the competing model of \citet{Davies_2016MNRAS}, in which reionization ends before $z\simeq6$ (blue/dashed curves in Fig. \ref{FIG:xHII_taues}). The defining feature is that $\avgMFP$ is a factor of 3 shorter than our fiducial model, resulting in large spatial variations in $\Gamma_{\rm HI}$ that are driven by galaxy clustering (and not neutral islands).   The original \citet{Davies_2016MNRAS} model did not include the effects of temperature fluctuations from reionization, which amounted to an implicit assumption that reionization ended early enough for the temperature-density relation to have relaxed to a tight power-law form \citep{Hui_1997MNRAS, McQuinnSanderbeck2016}.  Here we update the model by adding in temperature fluctuations consistent with reionization ending at $z\approx 6$.  We will denote this model with {\UrlFont early-reion-short-mfp}.\footnote{Note that what we call ``early reionization" here is still ``late" in the context of pre-Planck models. }  Lastly, we constructed a {\UrlFont late-reion-short-mfp} model that blends the two scenarios.   In this model (green/dot-dashed), the global neutral fractions are similar to our fiducial model at $z<6$, but $\avgMFP$ is a factor of 3 shorter.\footnote{{ Note that the mid-point of reionization differs somewhat between the {\UrlFont late-reion-short-mfp} and {\UrlFont late-reion-long-mfp} models, as shown in Fig. \ref{FIG:xHII_taues}.  This is because we set the ESMR maximum filter scale to the $\avgMFP(z=5.6)$ of the corresponding fluctuating UVB model, and then tune $\zeta$ to achieve the desired tail end of reionization. We emphasize, however, that, for the discussions in this paper, the most important differences between the models are in the global neutral fractions at $z<6$ and $\avgMFP$.}
}  Hence there are large \GammaHI\ fluctuations from both the short mean free path and the neutral islands.  

{ The right three columns of Table \ref{tab:models} show the mean values of \GammaHI\ in our models after matching mean fluxes to the observed values of \citet{Bosman_2018MNRAS}.  We have applied the correction factors in the appendix of \citet{D'Aloisio_2018MNRAS} in order to correct for finite resolution\footnote{\citet{D'Aloisio_2018MNRAS} quote the correction factors up to $z=5.8$.  We linearly extrapolate to obtain the correction factor at $z=6$. }.  These values are reasonably consistent with the measurements of \citet{D'Aloisio_2018MNRAS} given differences in the thermal histories of our simulations and in the mean fluxes measured by \citet{Bosman_2018MNRAS} and \citet{Becker_2015MNRAS}. }

{ In Fig. \ref{fig:taueff_lya} we compare our model distributions of \taueff\ against the measurements of \citet{Bosman_2018MNRAS} (gray shading) as well as the measurements of \cite{Eilers_2018ApJ} (orange) in 3 redshift bins.  Unless otherwise noted, we define \taueff\ in terms of an average of the continuum normalized flux, $F$, over 50$h^{-1}$ Mpc segments, \taueff$=-\ln\langle F\rangle$. } The shading spans the ``optimistic" and ``pessimistic" treatments of the non-detections in \citet{Bosman_2018MNRAS}.\footnote{Because we compare against these limits, we do not add noise to our synthetic skewers for Fig. \ref{fig:taueff_lya}.}  The former takes the $2\sigma$ upper limits on the fluxes for measurements and the latter assigns zero flux.


\subsection{Deconstructing the \lya forest fluctuations}
\label{sec:deconstructing}

\begin{figure}
\resizebox{8.2cm}{!}{\includegraphics{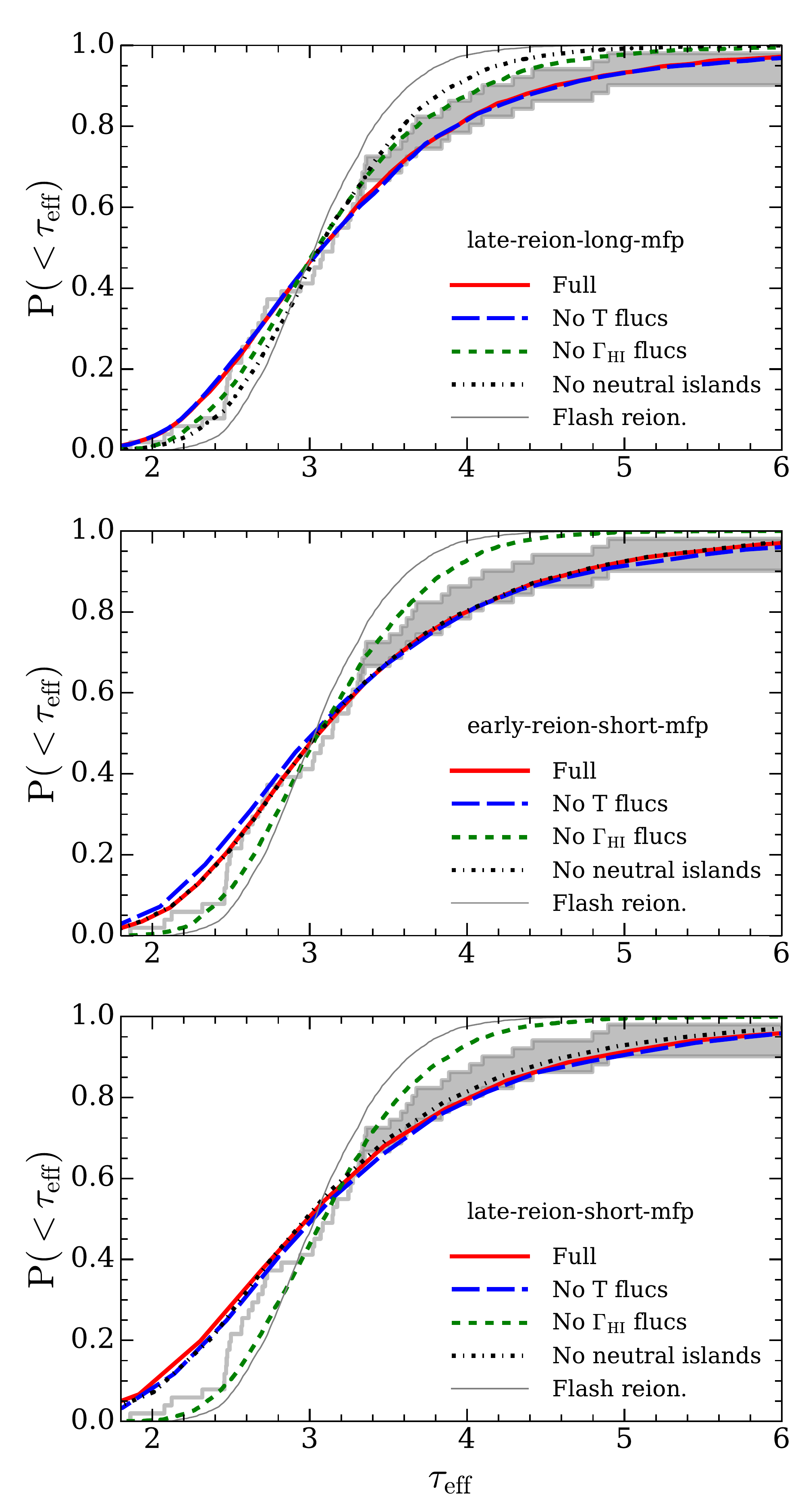}}
\caption{Deconstructing our models of the \lya forest opacity fluctuations.  Here we show a series of tests in which we leave out, one by one, the sources of fluctuation in our models.  ``No T flucs" denotes leaving out the temperature fluctuations from reionization, ``No $\Gamma_{\rm HI}$ flucs" denotes setting $\Gamma_{\rm HI}$ to a uniform value (outside of neutral regions). Lastly, ``No neutral islands" denotes removing the neutral islands, including their shadowing effects on the UVB.  Here we consider the distributions of \taueff\ at $z=5.6$.  These results show that, in the {\UrlFont late-reion-long-mfp} model, the opacity fluctuations are driven by neutral islands and their shadowing effects.}
\label{fig:deconstructing}
\end{figure}

\begin{figure}
\resizebox{8.2cm}{!}{\includegraphics{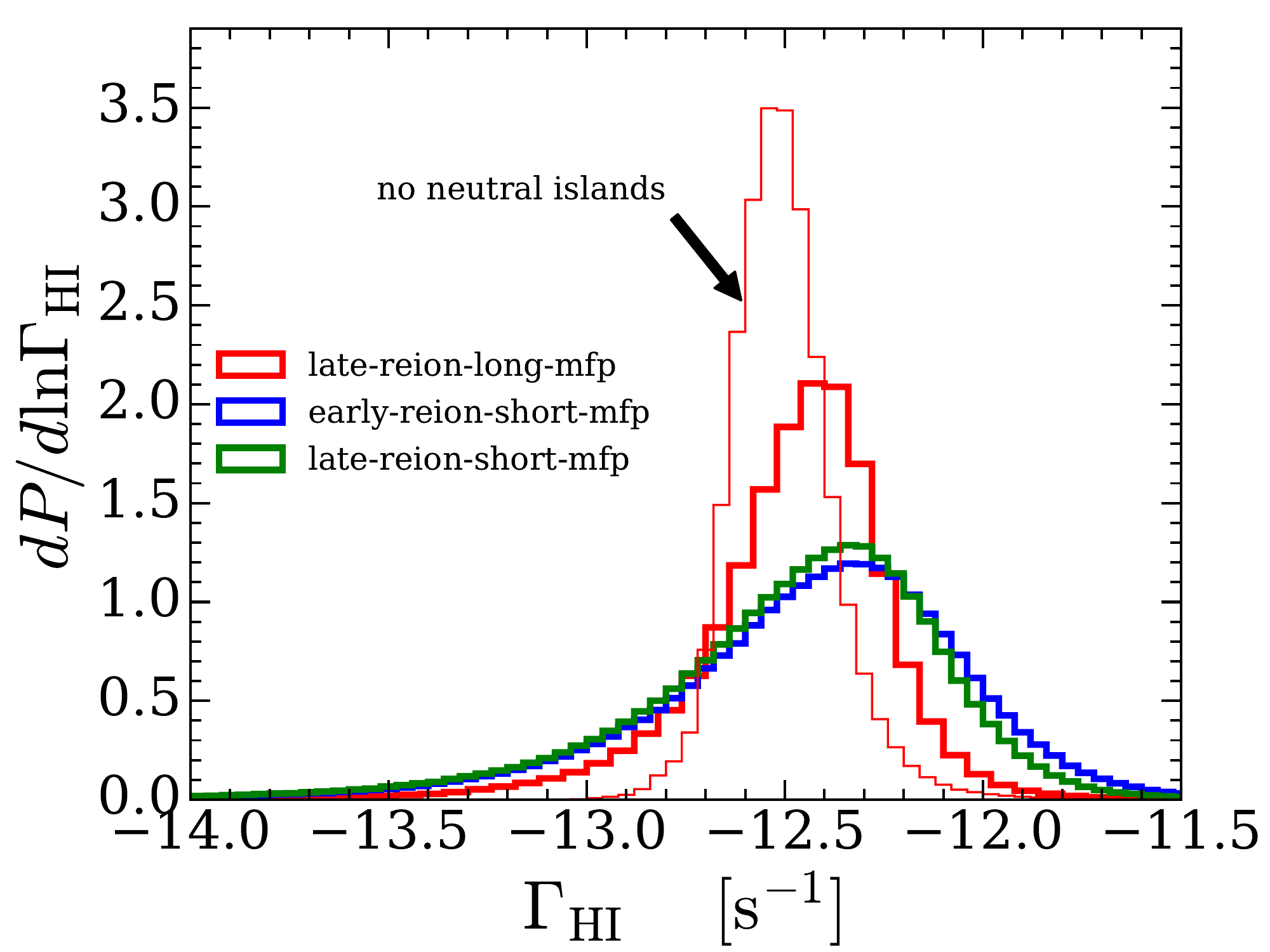}}
\caption{The probability distributions of $\Gamma_{\rm HI}$ in our models at $z=5.6$.  The thin red curve shows a variation of the {\UrlFont late-reion-long-mfp} model in which we leave out the neutral islands.  In this model, shadowing of the UVB by the neutral islands creates a low-$\Gamma_{\rm HI}$ tail associated with the highest Ly$\alpha$ forest opacities. }
\label{fig:gamflucs}
\end{figure}

One advantage of our piecemeal modeling is that we can straightforwardly explore the sources of \taueff\ fluctuation by simply leaving them out of the models.  In this section we deconstruct the opacity fluctuations to gain physical insight into their origin. 

We begin by visualizing the interplay between fluctuations. Fig. \ref{fig:vis} shows slices through the reionization redshift (left column), temperature (middle), and \HI\ photionization rate (right) fields at $z=5.6$.  The top, middle, and bottom rows correspond to the {\UrlFont late-reion-long-mfp}, {\UrlFont early-reion-short-mfp}, and {\UrlFont late-reion-short-mfp} models, respectively.  In both of the late scenarios there are large pockets of neutral hydrogen, tens of Mpc across, which can be seen as the black (white) regions in the temperature (photoionization rate) panels.  Note that the recently reionized gas adjacent to these pockets (closer to the I-fronts) is hotter than the gas elsewhere reionized long ago. The temperature field is more uniform in the {\UrlFont early-reion-short-mfp} model because the fluctuations have had more time to fade away.
Despite the bulk of reionization completing by $z=6$ in this model, there are still tiny neutral islands remaining because the short mean free path creates an extended tail of percent-level neutral fraction.  All three models exhibit large fluctuations in the UVB. To gauge the impact of the neutral islands on the UVB, the top-right panel of Fig. \ref{fig:vis} may be compared against the top-right panel of Fig. 2 in \citet{D'Aloisio_2018MNRAS}.  The latter shows what the UVB looks like with $\langle \mfp \rangle = 30 h^{-1}$ Mpc, but in the absence of neutral islands.        

Figure \ref{fig:deconstructing} shows the results of a series of tests where we leave out, one by one, the sources of fluctuation.  The curves labeled ``full" show the complete models. The curves labeled ``no $T$ flucs" correspond to neglecting the temperature fluctuations from reionization (we use the original temperatures from our hydro simulation, which exhibit a tight temperature-density relation).  Likewise, the curves labelled ``No \GammaHI\ flucs" correspond to adopting the nominal uniform \GammaHI\ of the hydro simulation everywhere outside of the neutral islands (within which \GammaHI$ =0$).  For the curves labeled ``No neutral islands," we remove the neutral islands as well as their shadowing effects on the UVB.\footnote{Here the temperature field is left the same, i.e. the gas is $10$ K where the neutral islands were.  However, we tested setting these regions to $10,000$ K instead and it did not change the results/conclusions reported in this section.}   Lastly, the curves labeled ``Flash reion." correspond to an instantaneous reionization at $z=7.5$.   Here we focus on the $z=5.6$ bin. 

Consider first {\UrlFont late-reion-long-mfp}, shown in the top panel of Fig. \ref{fig:deconstructing}.  Comparing the blue/long-dashed curve to the red/solid curve, we find that removing the temperature fluctuations has no noticeable impact on the width of the \taueff\ distribution.  In fact, this is true in all three models because the duration of reionization is too short to generate temperature fluctuations large enough to ``win" against the opposing effects of the \GammaHI\ fluctuations, at least when it comes to the \taueff\ fluctuations (see \citealt{D'Aloisio_2015ApJ} for a discussion on the role of reionization's duration in this context).  However, when they are present, we find that the temperature fluctuations do play a significant role in the statistics of the opacities, particularly in generating some of the most transmissive sight lines.  We will see one consequence of this in \S \ref{sec:LAEs}.  Foreshadowing, it appears that the interplay between temperature, \GammaHI, and neutral islands is non-trivial in these models.  We caution that no apparent effect on the \taueff\ distribution should not be interpreted as no effect at all.

 The green/short-dashed curve shows that setting $\Gamma_{\rm HI}$ to a uniform value outside of the neutral islands reduces the width and eliminates the high-opacity tail of the \taueff\ distribution.   We get a similar (but larger) effect if we instead remove the neutral islands entirely, which includes removing their shadowing effects on the UVB, as illustrated by the black/dot-dashed curve.  We find that the difference between the ``Full" and ``No \GammaHI\ flucs" models owes to the shadowing of the UVB in the vicinity of the neutral islands (which is lacking in the latter).  We conclude that both the neutral islands {\it and} their shadowing effects are critical for generating the high-opacity tail of the \taueff\ distribution in this model.  Note also that the ``No neutral island" distribution is wider than the ``Flash reion." distribution.  We find that this additional width owes to the temperature fluctuations from reionization.     

The situation is markedly different for the {\UrlFont early-reion-short-mfp} model (middle panel).  There is no noticeable difference in the \taueff\ distribution when we remove the neutral islands, as expected given the $0.5 \%$ global neutral fraction at $z=5.6$. In this model, the high-opacity tail is driven by large $\Gamma_{\rm HI}$ fluctuations whose origin lie in a short $\avgMFP$.  Lastly, we consider the {\UrlFont late-reion-short-mfp} model (bottom panel), which has a similar global neutral fraction to the {\UrlFont late-reion-long-mfp} model. Interestingly, removing the effects of neutral islands from this model has a relatively mild effect on the distribution.   The \GammaHI\ fluctuations are characteristically different between the {\UrlFont late-reion-long-mfp} and {\UrlFont late-reion-short-mfp} models because of the different values of $\avgMFP$.

These ideas are further elucidated in the top panel of Figure \ref{fig:gamflucs}, which compares the probability distributions of \GammaHI\ in our three models.  Note the wide range of \GammaHI\ that results from the short $\avgMFP$ in the {\UrlFont early-reion-short-mfp} model (blue).  This is similar to what is seen in the {\UrlFont late-reion-short-mfp} model (green).   The distribution has a different shape in the {\UrlFont late-reion-long-mfp} model (red).  The fluctuations are somewhat smaller, but there is a low-\GammaHI\ tail associated with the most opaque regions of the forest.   For reference, the thin red curve shows the distribution when we remove the shadowing effects of the neutral islands in that model. The low-\GammaHI\ tail disappears, illustrating the importance of the shadowing for generating extended low-opacity regions.

\subsection{Long \lya Troughs}
\label{sec:troughs}

\begin{figure}
\resizebox{8.4cm}{!}{\includegraphics{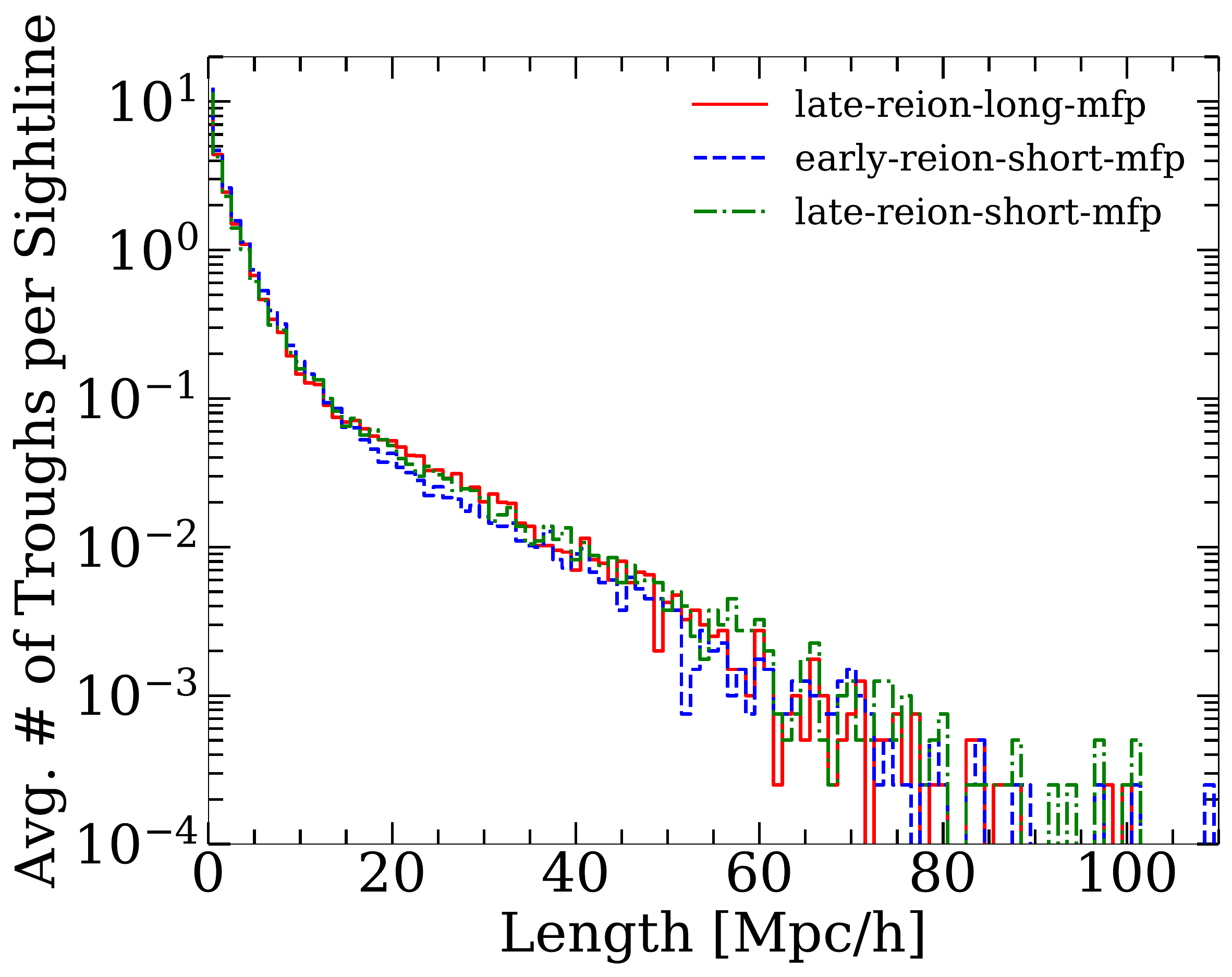}}
\caption{ The $5.5 < z < 5.9$ \lya\ trough size distributions in our models. Here we show the average number of troughs of a given size per sight line. Note that this quantity yields the frequency of troughs when the average number is much less than unity.  The simulated sight lines are processed with ${\rm S/N}=150$ per pixel.}
\label{fig:trough_examples_count}
\end{figure}

\begin{figure*}
\resizebox{16cm}{!}{\includegraphics{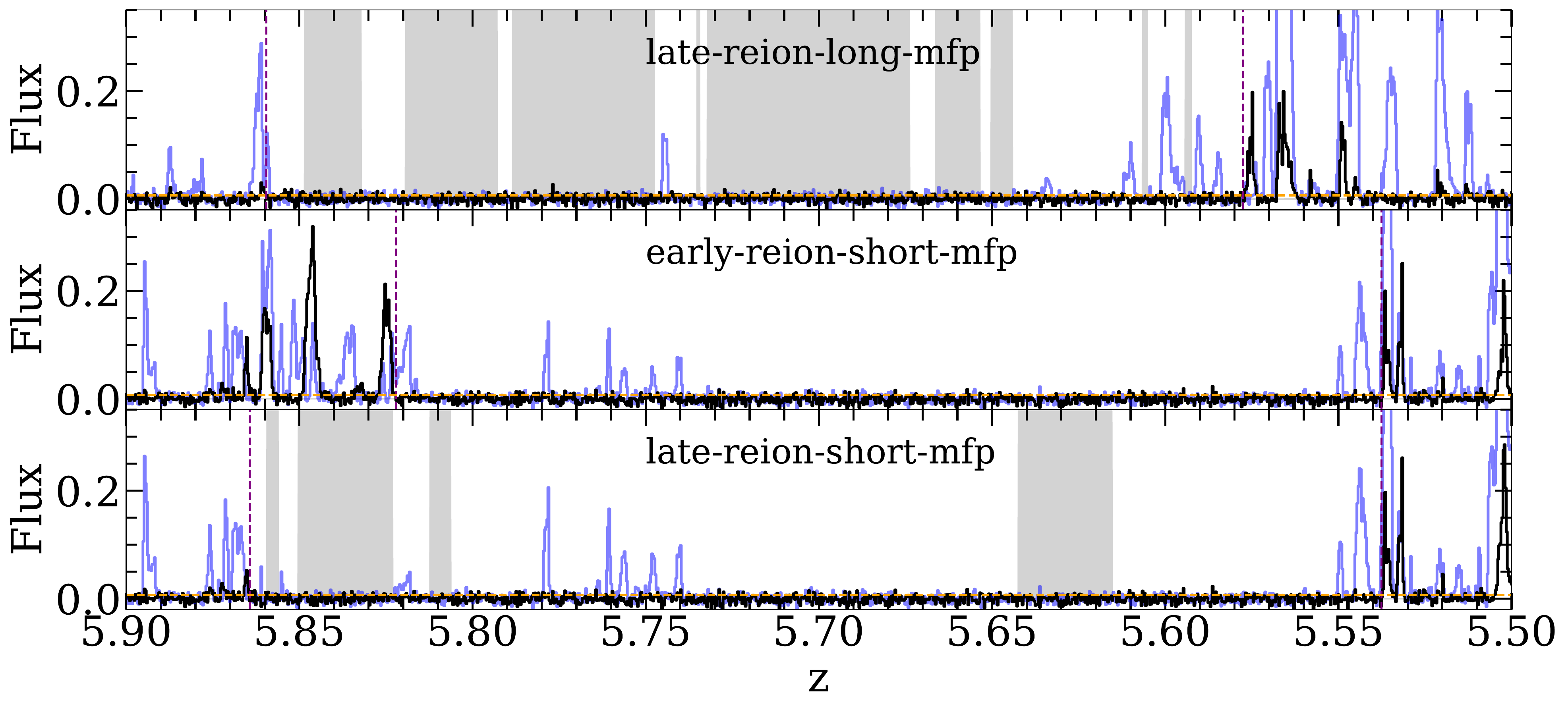}}
\caption{Examples of long Ly$\alpha$ troughs with lengths and Ly$\beta$ opacities similar to those of the J0148 trough observed by \citet{Becker_2015MNRAS}. Ly$\alpha$ and Ly$\beta$ transmission are shown in black and blue, respectively. Vertical dashed lines mark the extents of the Ly$\alpha$ troughs. From top to bottom, troughs lengths are 100, 109, 101\mpc\ respectively. The Ly$\beta$ \taueff\ are 5.213, 5.137, and 5.152.  In all of our models, the most extreme Ly$\alpha$ troughs occur in cosmological voids.  Only in the {\UrlFont late-reion-long-mfp} model are the neutral islands physically necessary for creating the troughs. { We note that the most extreme troughs originate from different sight lines in different models.  As a result, each example depicts a different sight line (with different density structure).}}
\label{fig:trough_examples}
\end{figure*}

In this section we examine whether our models reproduce long Ly$\alpha$ troughs like the one observed in J0148.  We focus on the redshift range $5.5 < z < 5.9$ spanned by the J0148 trough.  Because of the steep evolution in \GammaHI\ over this interval \citep{D'Aloisio_2018MNRAS}, we include redshift evolution in the manner described in \S \ref{sec:methods_skewers}.  We also add noise with ${\rm S/N}=150$ per pixel, approximately matching the noise characteristics of the J0148 spectrum in \citet{Becker_2015MNRAS}.  We identify \lya troughs as regions between ``peaks" where the continuum normalized flux is above 2$\sigma$ of noise at four consecutive pixels \citep{Becker_2015MNRAS, Keating_2019arXiv}. 

In Fig. \ref{fig:trough_examples_count}, we show the  \lya trough size distribution extracted at $5.5<z<5.9$. Despite the physical differences between models noted in \S \ref{sec:deconstructing}, the distributions are very similar, with the longest troughs extending to $\approx 110h^{-1}$ Mpc, similar to the length of the J0148 trough. The incidence of $\sim 100$ Mpc troughs, however, is quite low, occurring in $1-2$ out of 4000 sight lines in all three models.  These rates are much lower than the roughly $\gtrsim 1/100$ rate inferred from the existence of J0148, but we note that this discrepancy likely owes to our small box size of $L=200 h^{-1}$ Mpc.  Capturing the incidence rates of these long troughs requires much larger volumes.

In Fig.~\ref{fig:trough_examples}, we show one example trough from each model, selected with similar characteristics to the J0148 trough. { We note that the most extreme troughs originate from different sight lines in different models.  As a result, each example depicts a different sight line (with different density structure).} The length of these troughs are $\gtrsim100 h^{-1}$ Mpc and they are more dark (\taueff$>7.7$) than in J0148. One of the defining features of the J0148 trough is its significant \lyb transmission.  The troughs in Fig. \ref{fig:trough_examples} were selected to have similar \lyb transmission (\taueff$\simeq5.2$).  The vertical dashed lines in the figure mark the ends of the troughs, and the gray bands span the locations of neutral islands. 

The trough we display here for the {\UrlFont early-reion-short-mfp} model does not intersect any neutral islands, illustrating that they are not necessary for generating long troughs in this model.  However, we find that, even in this model, with $x_{\rm HI}(z=5.8) = 0.01$,   the most extreme troughs more often do contain neutral islands. For example, we find that $6/9$ troughs with lengths $> 80 h^{-1}$ Mpc contain neutral islands.  If there is even a tiny amount of neutral hydrogen left in the IGM, the longest Ly$\alpha$ troughs are likely to intersect it.  However, we emphasize that the neutral islands are physically necessary for generating these troughs only in the {\UrlFont late-reion-long-mfp} model. We find that the trough size distribution, and even the individual troughs by inspection, look nearly identical in the ``No neutral islands" version of the {\UrlFont early-reion-short-mfp} model.   

In summary, all of our models successfully produce $5.5 < z < 5.9$ Ly $\alpha$ troughs that are reasonably similar in character to the J0148 trough, albeit at much lower incidence rates owing to our finite box size.


\section{Testing the Models}

\label{sec:testing}

In this section we search for ways that the three scenarios can be tested and distinguished.  As we will see, the models appear quite similar in their \lya/\lyb forest statistics. 

\subsection{Thermal history of the IGM}
\label{sec:thermal_hist}

\begin{figure}
\resizebox{8.4cm}{!}{\includegraphics{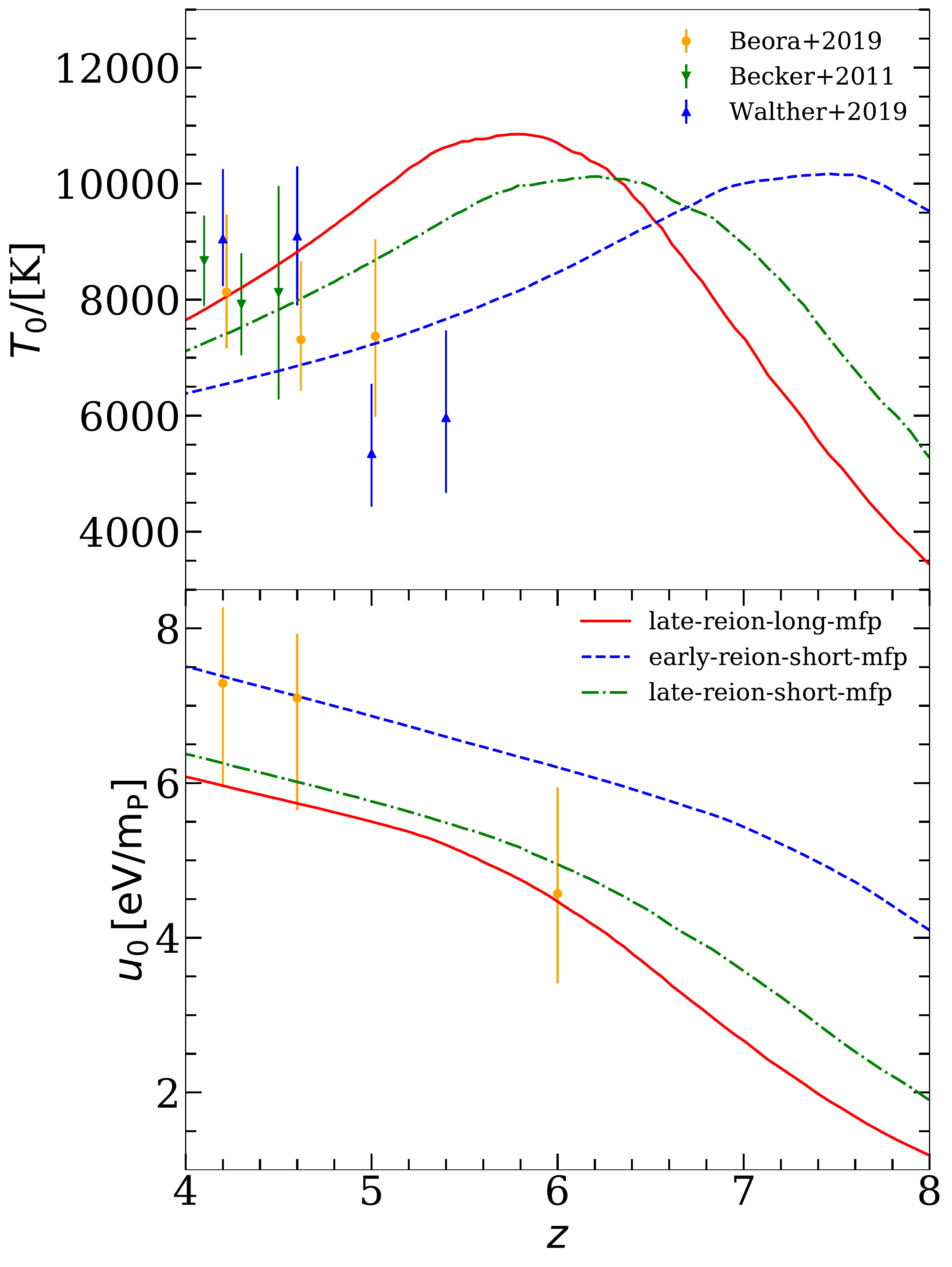}}
\caption{Thermal histories in our three reionization models. {\it Top:} Evolution of the gas temperature at mean density of the Universe (\to).  {\it Bottom:} Cumulative heating per baryon at the mean density (\uo), which parameterizes the degree of pressure smoothing of the gas.  The data points correspond to a selection of recent Ly$\alpha$ forest measurements.  Note that our models do not include heating from \HeII\ reionization, which is expected to begin playing a significant role at $z<5$.}
\label{fig:thermal_hist}
\end{figure}

\begin{figure}
\resizebox{8.4cm}{!}{\includegraphics{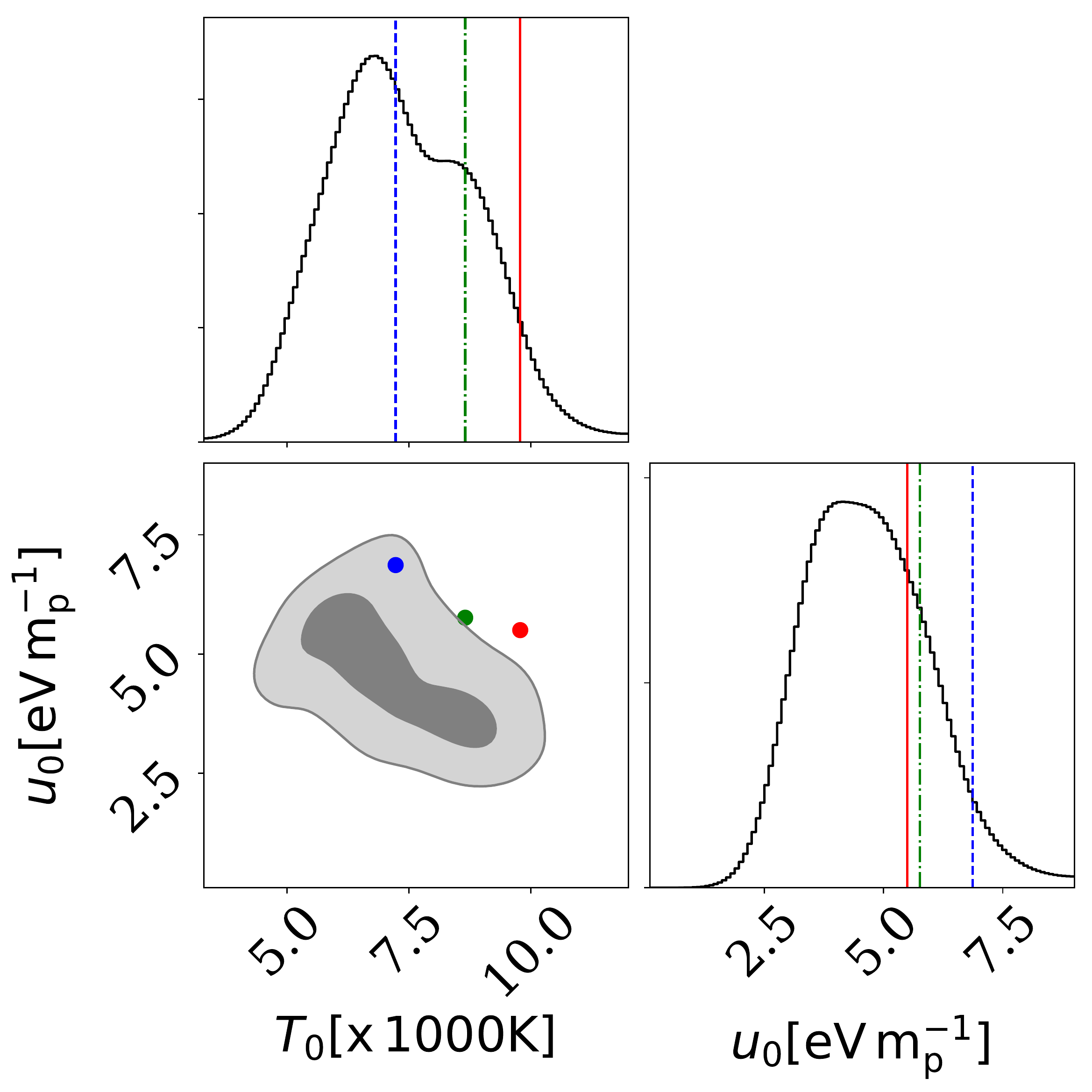}}
\caption{Posterior distributions for the parameters \to\ and \uo\ from \citet{Boera_2019ApJ}, obtained from flux power spectrum measurements at $z=5$. The red, blue, and green dots correspond to the {\UrlFont late-reion-long-mfp}, {\UrlFont early-reion-short-mfp}, and {\UrlFont late-reion-short-mfp} models, respectively 
The grey contours show the 68 and 95 percent credibility regions. The black histograms display the one-dimensional marginalized posterior distributions for the parameters. Note that the evolution of \to\ and \uo\ are shown in Fig~\ref{fig:thermal_hist}.}
\label{fig:thermal_hist_cont}
\end{figure}

As the timing of reionization is different between the ``late" and ``early" models, a natural starting place is the thermal history of the IGM.  Recent studies have pushed \lya forest temperature measurements up to $z\approx 5.4$.  These measurements would probe temperatures near overlap in the late reionization scenarios, making them an important consistency check.  The top panel of Fig. \ref{fig:thermal_hist} shows the IGM temperature at the cosmic mean density, \to, in our three models. For simplicity, we neglect photoheating from \HeII ionization because (1) our focus is on $z\sim5$ temperature measurements, and (2) observational constraints suggest that the heating from \HeII\ reionization likely does not pick up until after $z\sim5$ \citep[see e.g.][]{2000MNRAS.318..817S,Becker_2011MNRAS,Walther_2019ApJ, Puchwein_2015MNRAS, Sanderbeck_2016MNRAS, D'Aloisio_2017MNRAS, 2017ApJ...841...87L}.  We compare our calculations against some of the most recent high-$z$ measurements.    

The two late scenarios place the temperature bump, which occurs near the end of reionization, at $z= 5.5-6.0$.  In these cases we should start to see a rise in temperatures as measurements are pushed towards $z\approx 5.5$. There is, however, no evidence for this rise as of yet.  In fact, taken at face value, the measurements disfavor the late reionization models.  However, this conclusion comes with two caveats: (1) In formulating our thermal histories, we have adopted the reionization heating calculations of \citet{D'Aloisio_2019ApJ} based on the I-front speeds in our ESMR reionization models.  It is possible that the actual speeds were significantly slower due to the poorly understood role of self-shielded absorbers, which are not captured in our ESMR models (and are likely also not captured in most cosmological radiative transfer simulations).  This could have resulted in lower temperatures, particularly near the end of reionization.  However, the reduction in I-front speeds would have to be dramatic. To put some numbers to this, lowering I-front speeds by a factor of 20 from $10^4$ to $5\times10^2$ proper km/s reduces $T_{\mathrm{reion}}$ from $\approx 27,000K$ to $17,000 K$ \citep{D'Aloisio_2019ApJ}; (2) The state-of-the-art temperature measurements rely on forward-modeling with hydro simulations that do not capture the \GammaHI\ and temperature fluctuations, and inhomogeneous pressure smoothing, from reionization.  This could potentially impact the measurements in at least two ways.  Firstly, the fluctuations may effectively down-weight the contribution of cold or hot regions to the flux power spectrum, biasing the extracted temperatures. However, previous studies suggest that these effects are likely subtle \citep{Onorbe_2019MNRAS,2019arXiv190704860W}. Secondly, the 1D flux power spectra in early and late reionization models are known to be degenerate; a warm, recently reionized IGM is difficult to distinguish from a cold, pressure-smoothed IGM \citep[see e.g.][]{2019arXiv190704860W}.  The latest measurements account for this degeneracy, but it would be prudent to revisit this topic with more realistic simulations, especially as larger observational data sets are being acquired.

The Ly$\alpha$ forest is sensitive not only to the instantaneous temperature (by Doppler broadening), but also to the reionization thermal history through pressure smoothing of the gas.  Recent studies have made progress in disentangling the pressure smoothing effects, which, even at lower redshifts, can be used to constrain reionization in principle.  \citet{Walther_2019ApJ} parameterized the smoothing with a length scale (termed the pressure smoothing scale) that characterizes these effects in the 1D flux power spectrum.    Alternatively, \citet{Boera_2019ApJ} casted these effects into the cumulative heat per proton injected into gas at the mean density, $u_0$ \citep{Nasir_2016MNRAS}.  Our mock Ly$\alpha$ forest spectra cannot reliably model the small-scale flux power spectrum because our hydro simulation does not have sufficient resolution.  Equally important, the pressure smoothing in our models is unrealistic because the inhomogeneous effects of reionization are added in post-processing. We can, however, still get a handle on the pressure smoothing differences between our models  by calculating $u_0$. 

We calculate $u_0$ for each mean-density gas parcel in our simulations using that 
\begin{equation}
u_{0} = u_{\rm ini,0}+ \int_{z_{\rm{o}}}^{z_{\rm{re}}} \frac{\mathscr{H}}{\bar{\rho}}\frac{dz}{H(1+z)},
\label{eq:uo}
\end{equation}
where $\,u_{\rm ini,0}\,$ corresponds to the impulsive heat injection by an I-front at $z=$\zre, and the second term corresponds to the post-reionization photoheating.  In the latter, $\bar{\rho}={\rho}_{\rm{crit}}{\Omega}_{\rm{b}}{(1+z)}^{3}$ is the mean baryonic matter density, and $\mathscr{H}$ is the net photoheating rate per unit volume (here for \HI\ and \HeI).\footnote{Note that $\mathscr{H}$ is a function of temperature through the recombination rates.  For a given gas parcel, we use the temperature obtained from Eq.~(\ref{eq:temp}) to calculate \uo at a given redshift.}   If the temperature to which the I-front heats the gas is $T_{\rm reion}$, then we take $u_{\rm ini,0} = (3/2) k_{B}T_{\rm reion}/\mu m_H$, where we adopt the mean molecular mass of $\mu=0.61$ for ionized hydrogen.  Gas parcels are considered to be mean-density if they fall within 2\% of $\Delta=1$ and the \uo\ that we quote here is averaged over this bin. The bottom panel of Fig. \ref{fig:thermal_hist} compares our models of \uo\ with measurements by \citet{Boera_2019ApJ}. The post-reionization \uo\ in {\UrlFont early-reion-short-mfp} is $\sim 25$ \% higher than in the late models owing to the earlier start of reionization.  Figure \ref{fig:thermal_hist_cont} compares our models to the $z=5$ joint posterior distribution of $T_0$ and $u_0$ from \citet{Boera_2019ApJ}.  The contours correspond to 68 and 95 \% credibility regions and the red, blue, and green dots correspond to the {\UrlFont late-reion-long-mfp}, {\UrlFont early-reion-short-mfp}, and {\UrlFont late-reion-short-mfp} models, respectively.  We note that the models roughly trace the degeneracy direction of the measurements.

In summary, current thermal history measurements do not conclusively favor any our reionization models due to large uncertainties.  However, future measurements of the temperature at $z=5.5$ and of the pressure smoothing will likely play an important role in distinguishing between scenarios.

\subsection{\lya/\lyb forest statistics}\label{sec:peak}

\begin{figure*}
\begin{center}
\resizebox{16cm}{!}{\includegraphics{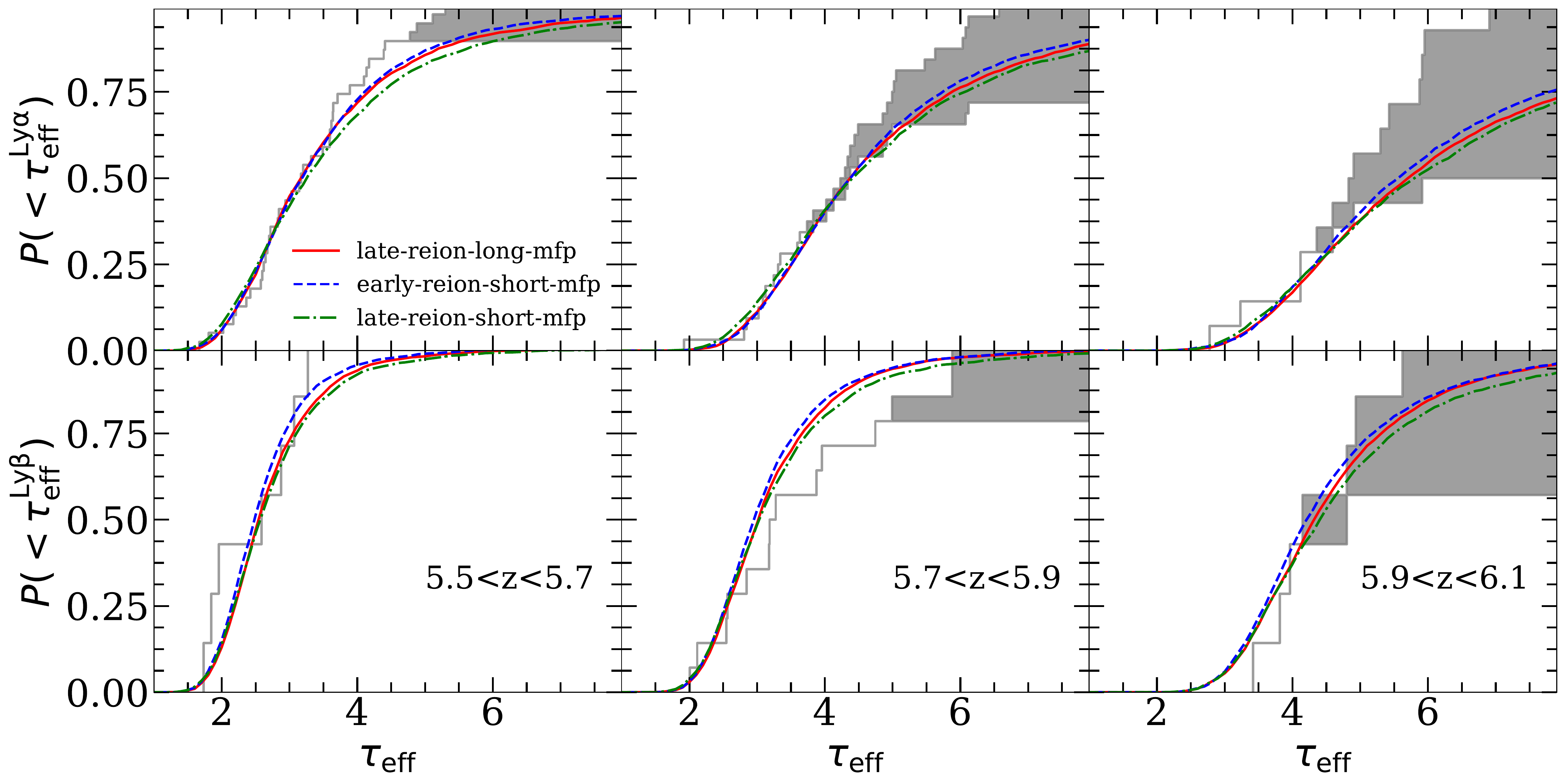}}
\hspace{-0.28cm}
\end{center}
\caption{A comparison of Ly$\alpha$ (top) and Ly$\beta$ (bottom) \taueff\ cumulative distributions.  The grey histograms show the measurements of \citet{Eilers_2019ApJ}, where the shading spans the ``optimistic'' and ``pessimistic'' limits (see text in \S \ref{sec:scenarios} for a description). Following the convention of \citet{Eilers_2019ApJ}, we define \taueff\ by an average over 27 \mpc\ segments (as opposed to the $50h^{-1}$ Mpc adopted in the rest of this paper).  Our models yield very similar distributions, despite having different global neutral fractions at these redshifts. }
\label{fig:taueff_cdf}
\end{figure*}

\begin{figure*}
\begin{center}
\resizebox{14cm}{!}{\includegraphics{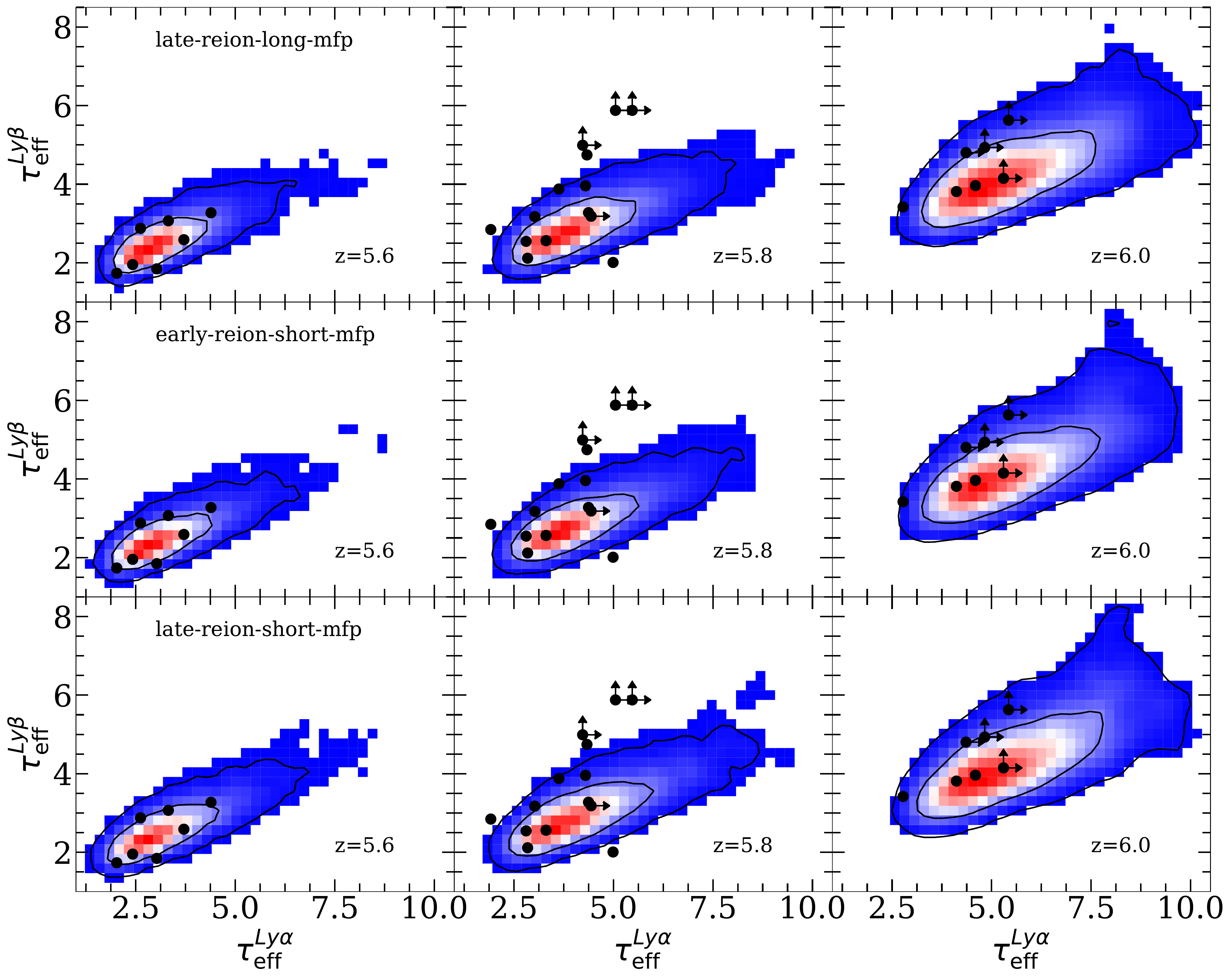}}
\hspace{-0.28cm}
\end{center}
\caption{Correlation between Ly$\alpha$ and Ly$\beta$ effective optical depths.  Here we define \taueff\ the same way as in Fig. \ref{fig:taueff_cdf}.  The contours correspond to 68 and 95 percent regions and the data points show the observational measurements of \citet{Eilers_2019ApJ}.  The rows show different models while the columns correspond to 3 redshift bins.  The models are broadly similar in this correlation.   }
\label{fig:lya_lyb_correlation}
\end{figure*}

\begin{figure*}
\resizebox{14cm}{!}{\includegraphics{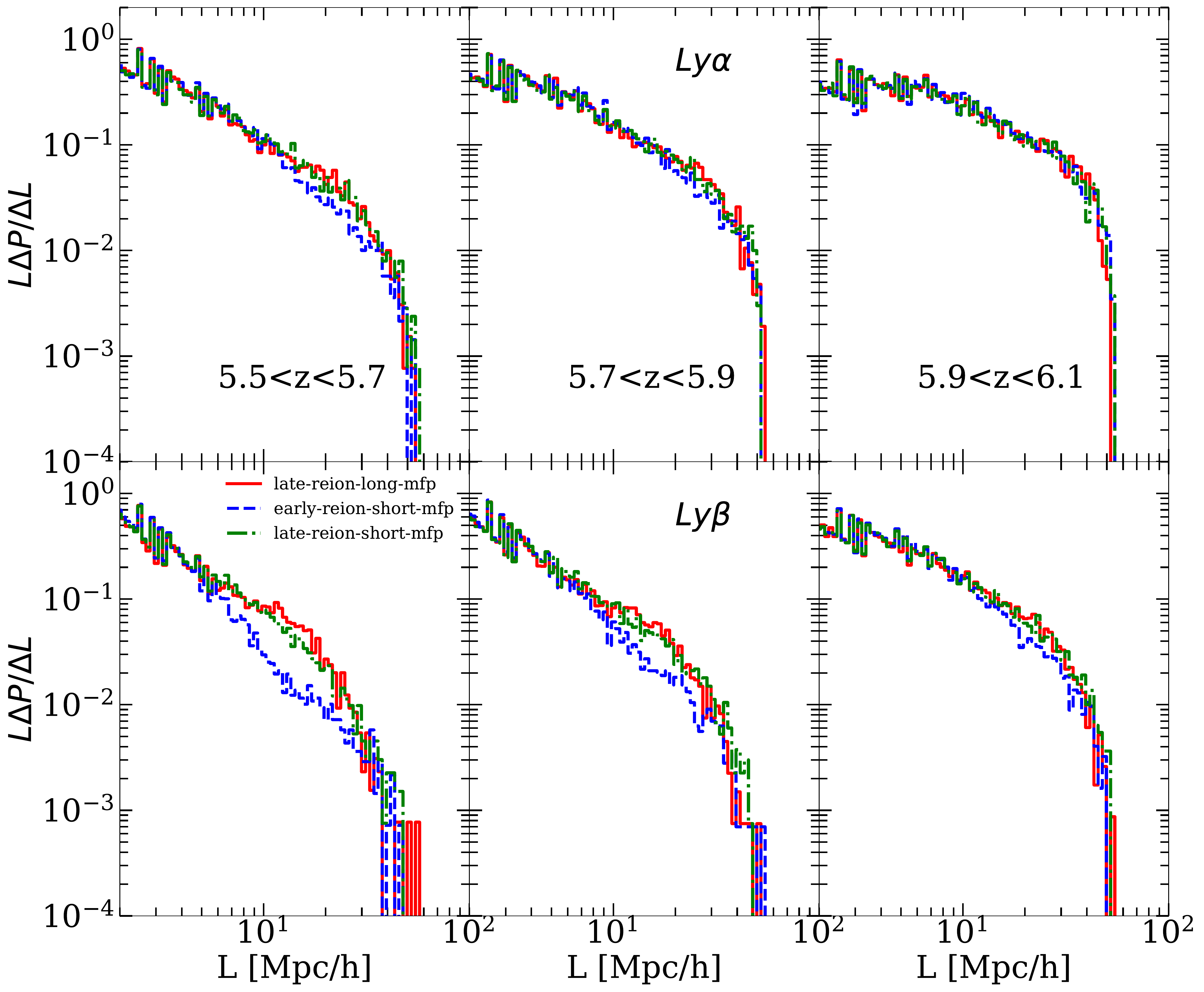}}
\caption{ Trough size distributions (or ``dark gap distributions") for Ly$\alpha$ (top) and Ly$\beta$ (bottom).  In Ly$\alpha$ the distributions are indistinguishable, but in Ly$\beta$ the {\UrlFont early-reion-short-mfp} model shows fewer troughs of intermediate size ($5 \lesssim L \lesssim20$ \mpc), especially in the $z=5.6$ bin.}
\label{fig:dark_gaps}
\end{figure*}

\begin{figure}
\resizebox{8.5cm}{!}{\includegraphics{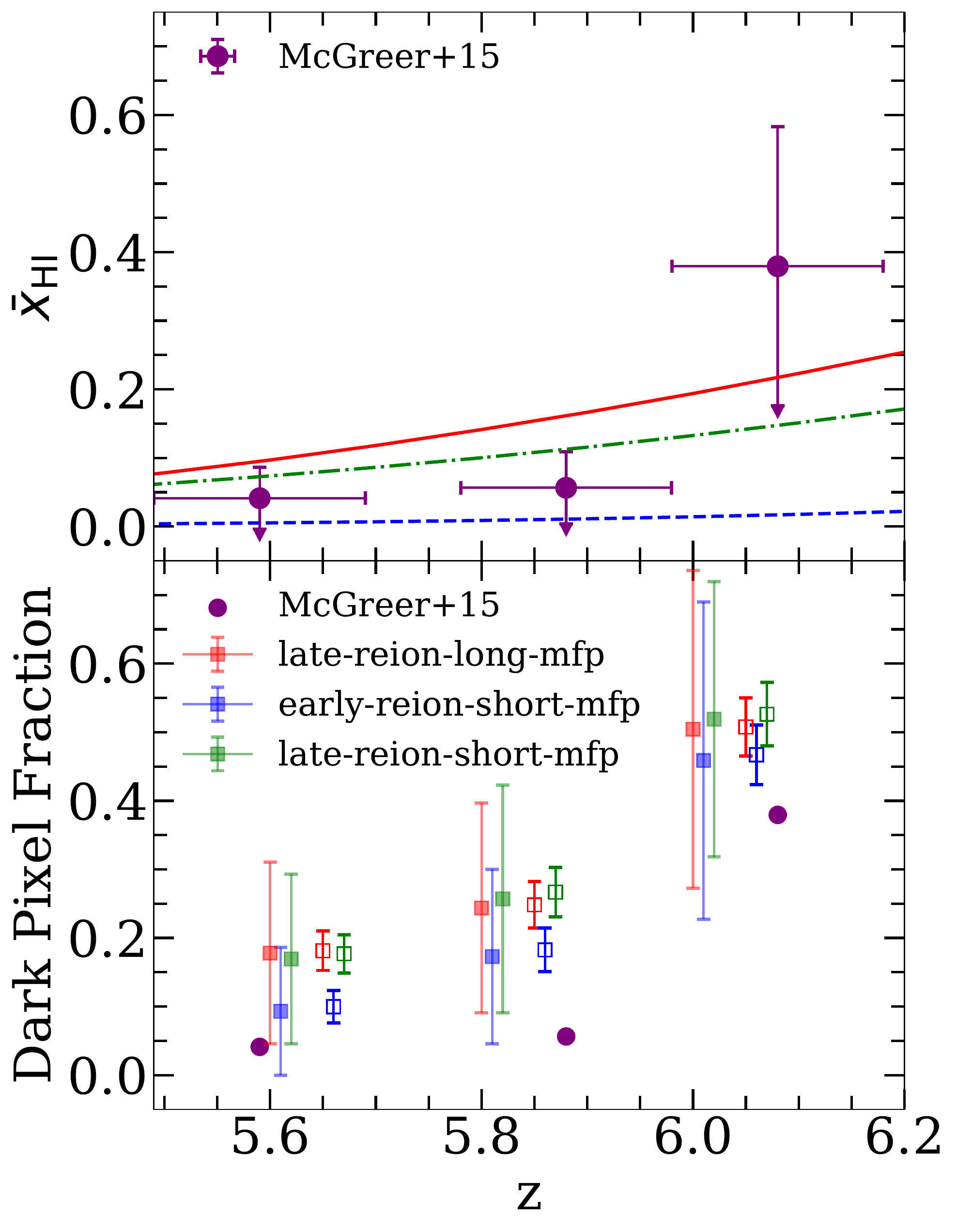}}
\caption{ { The evolution of the global \HI\ fraction using the dark pixel fraction.   {\it Top panel:} The reionization histories of our models compared against 1$\sigma$ upper limits on the neutral fraction measured by \citet{McGreer_2015MNRAS}. The upper limits were obtained from dark pixels in the coeval Ly$\alpha$ and Ly$\beta$ forests of their ``best sample."  {\it Bottom panel:}  The dark pixel fractions in our mock Ly$\alpha$ and coeval Ly$\beta$ forests. We have matched (approximately) our model spectra to the characteristics of the best sample in \citet{McGreer_2015MNRAS}, and we have used their procedure for obtaining \HI\ fractions.  The data points correspond to median values and the error bars display 10th and 90th percentiles, estimated by bootstrap resampling.  The closed and open symbols (offset for clear presentation) correspond to using 4 and 100 sight lines per redshift bin, respectively, where the latter corresponds to futuristic measurements.  The purple circles show the measurements of \citet{McGreer_2015MNRAS}. }}
\label{fig:dark_pixels}
\end{figure}

\subsubsection{Ly$\beta$ opacities}

We now turn to \lya and \lyb forest statistics. We begin with the cumulative distribution of \taueff\ in Ly$\beta$.   The Ly$\beta$ opacities are a useful window into the fluctuations because they probe different temperatures and densities.   Recently, \citet{Eilers_2019ApJ} measured this distribution using a sample of 19 quasar sightlines between $5.5 \lesssim z \lesssim 6.1$.  Particularly at $z\gtrsim 5.8$, they found that fluctuating UVB and temperature models, akin to those of \citet{Davies_2016MNRAS} and \citet{D'Aloisio_2015ApJ},  under-predict the Ly$\beta$ opacities when they are tuned to match the observed distribution of Ly$\alpha$ opacities. It is interesting to explore this discrepancy further with our models, since they contain neutral islands that can alter the correlation between Ly$\alpha$ and Ly$\beta$ opacities.  

In this section we follow the convention of \citet{Eilers_2019ApJ} in which we measure \taueff\ along segments of length $27 h^{-1}$ Mpc, rather than $50 h^{-1}$ Mpc.   The UVB in our models are iteratively rescaled in the ionized regions to match the Ly$\alpha$ mean fluxes of \citet{Eilers_2019ApJ}, excluding non-detections, and for Ly$\beta$ we include foreground Ly$\alpha$ absorption. To each Ly$\beta$ skewer we add opacities from a randomly chosen skewer from our hydro simulation at $z \approx  (1+z_{Ly\alpha})  \lambda_{Ly\beta}/\lambda_{Ly\alpha} - 1$, where $\lambda_{Ly\beta}$ and $\lambda_{Ly\alpha}$ are the rest-frame Ly$\alpha$ and Ly$\beta$ wavelengths, respectively, and $z_{Ly\alpha}$ is the redshift of interest in the Ly$\alpha$ forest. We rescale the foreground hydro skewers to match the mean fluxes reported by \citet{Irsic_2017PhRvD}.

Fig~\ref{fig:taueff_cdf} shows a comparison of \taueff\ cumulative distributions for Ly$\alpha$ (top row) and Ly$\beta$ (bottom row) with the measurements of \citet{Eilers_2019ApJ}.\footnote{  \citet{Eilers_2019ApJ} noted the importance of modelling noise when comparing models to observed distributions, particularly for the high-opacity tails of the distributions.  However, just as in Fig. \ref{fig:taueff_lya}, we do not add noise to our mock spectra when calculating our distributions here because we are comparing against the optimistic and pessimistic limits of the measurements. }  All of our models provide reasonable matches to the measurements, with the exception of the $5.7 < z < 5.9$ bin, where the models do not accommodate the highest Ly$\beta$ opacities observed. Examining the Ly$\beta$ distributions in the 2nd row, the two late reionization models are systematically more opaque than the early model (blue/dashed), but the difference is rather subtle.

We have also examined the correlation between Ly$\alpha$ and Ly$\beta$ \taueff, as shown in Fig. \ref{fig:lya_lyb_correlation}.  The contours correspond to 68 and 95 \% levels.  For this plot, we added noise with $S/N = 150$  to roughly match the typical values in \citet{Eilers_2019ApJ}, but we have checked that our conclusion on the similarity of the models is unaffected by the particular choice.  Our main conclusions here are: (1) all of the models provide reasonable matches to the data, { but they struggle to accommodate the largest Ly$\beta$ opacities in the $z=5.8$ bin}; (2) Unfortunately, the models are very similar in these statistics, and we do not find any obvious differences that can help in distinguishing between them.

\subsubsection{Trough size distributions}

In \S \ref{sec:troughs} we showed that our models have very similar \lya trough-size distributions. We now examine the trough distributions in Ly$\beta$.  It is worth noting that our simulations are likely not converged on the prevalence of transmission peaks, given their sensitivity to fluctuations on extremely small scales.  However, we emphasize that our goal here is to search for any obvious qualitative differences between the models.  Future models suitable for comparison against data will have to address numerical convergence as well as the in-homogeneity of data sets in spectral resolution and quality.     

In Fig.~\ref{fig:dark_gaps}, we show the evolution of size distributions in Ly$\alpha$ (top row) and Ly$\beta$ (bottom row). The troughs are extracted using same method as discussed in \S \ref{sec:troughs}, and we also include the effects of evolution in \GammaHI\ over the $\Delta z = 0.2$ redshift bins.\footnote{Note that the binning limits the length of the longest \lya/\lyb trough.}   As in \S \ref{sec:troughs}, the Ly$\alpha$ size distributions are nearly identical.  There are some noticeable differences, however, in the Ly$\beta$ distributions. Specifically, the {\UrlFont early-reion-short-mfp} model (blue/dashed) exhibits a relative deficit of troughs at lengths $5 \lesssim L \lesssim20$ \mpc, particularly in the $5.5<z<5.7$ bin.   The deficit of intermediate-size troughs reflects their different physical origin compared to the late models.  In the former, the intermediate troughs correspond to highly ionized regions with very low \GammaHI, whereas in the latter they typically contain extended islands of neutral hydrogen.   These differences become more difficult to discern towards higher redshifts as the forest becomes more saturated.

\subsubsection{Dark pixel fraction}

Lastly, we examine the forest dark pixel fraction, which has been used to place model independent limits on the global neutral fraction at $z \lesssim 6$ \citep{McGreer_2015MNRAS}. Given the difference in neutral fraction at $z\sim 6$ between the late and early scenarios, the dark pixel fraction seems like a natural possibility for distinguishing the models. 

{ The top panel of Fig.~\ref{fig:dark_pixels} compares the 1$\sigma$ upper limits on the neutral fraction of \citet{McGreer_2015MNRAS} against the reionization histories in our models. Another way to compare our models to the measurements is to forward model (to the extent possible) their forest observations and extract an ensemble of mock dark fraction measurements. The distribution of these mock measurements can then be compared against the actual measurements.}  For this calculation, we rebin to $3.3$ Mpc pixels as in \citet{McGreer_2015MNRAS} and we add a Gaussian noise with ${\rm S/N}=e^{5.5}$ per pixel to mimic the average signal-to-noise of their ``best sample."  We measure the dark pixel fraction by counting the pixels with negative flux in both \lya and \lyb simultaneously, and then multiply this number by four.  This accounts for the fact that the noise should create upward and downward excursions with equal probability.  We use 4 sight lines per redshift bin, roughly similar to the number of sight lines with both Ly$\alpha$ and Ly$\beta$ coverage in the best sample in \citet{McGreer_2015MNRAS}.  { We generate a distribution of mock measurements using $1000$ bootstrap samples.}  We also explore a ``futuristic" scenario assuming 100 sight lines per bin.

{   In the bottom panel of Fig.~\ref{fig:dark_pixels}, we compare the mock dark pixel fraction distributions to the measurements from \citet{McGreer_2015MNRAS} (circles).  The squares show median values from our models while the error bars denote 10th and 90th percentiles. The filled and open squares correspond to the current and futuristic scenarios, respectively, where the latter is offset along the x-axis for clarity.}  The dark pixel fraction is systematically lower in the {\UrlFont early-reion-short-mfp} model, but only modestly. In fact, the models are again similar in this statistic across all three redshifts, particularly at $z=6$, where one might hope to have the best chance of discerning them.  These results suggest that it may be difficult to discern early and late reionization scenarios with dark pixel statistics, even in a futuristic analysis with 100 sight lines per redshift bin, owing to saturation of the forest for ionized regions with low \GammaHI\ in the former.   { On the other hand, our results suggest that future measurements of the dark pixel fraction could rule out all three models. We note a mild tension between our models and the current measurements, evident in the two lowest redshift bins (compare filled squares with circles).}

In summary, we have examined a few \lya\ and \lyb\ forest statistics in search of an easy discriminator for our three distinct reionization scenarios, which have all been tuned to match the mean opacity and scatter in the $z>5.5$ Ly$\alpha$ forest. We found that the models are broadly similar in the statistics that we examined, with the exception of the frequency of intermediate-size ($5 \lesssim L \lesssim 20$ \mpc) troughs in the Ly$\beta$ forest.

\subsection{\lya emitters}
\label{sec:LAEs}

LAEs have emerged as a useful window into the nature of the high-$z$ forest opacity fluctuations \citep{2018ApJ...860..155D, Becker_2018ApJ}.   In this section we explore the prospect of distinguishing our models with LAE surveys.

\begin{figure*}
\begin{center}
\resizebox{17cm}{!}{\includegraphics{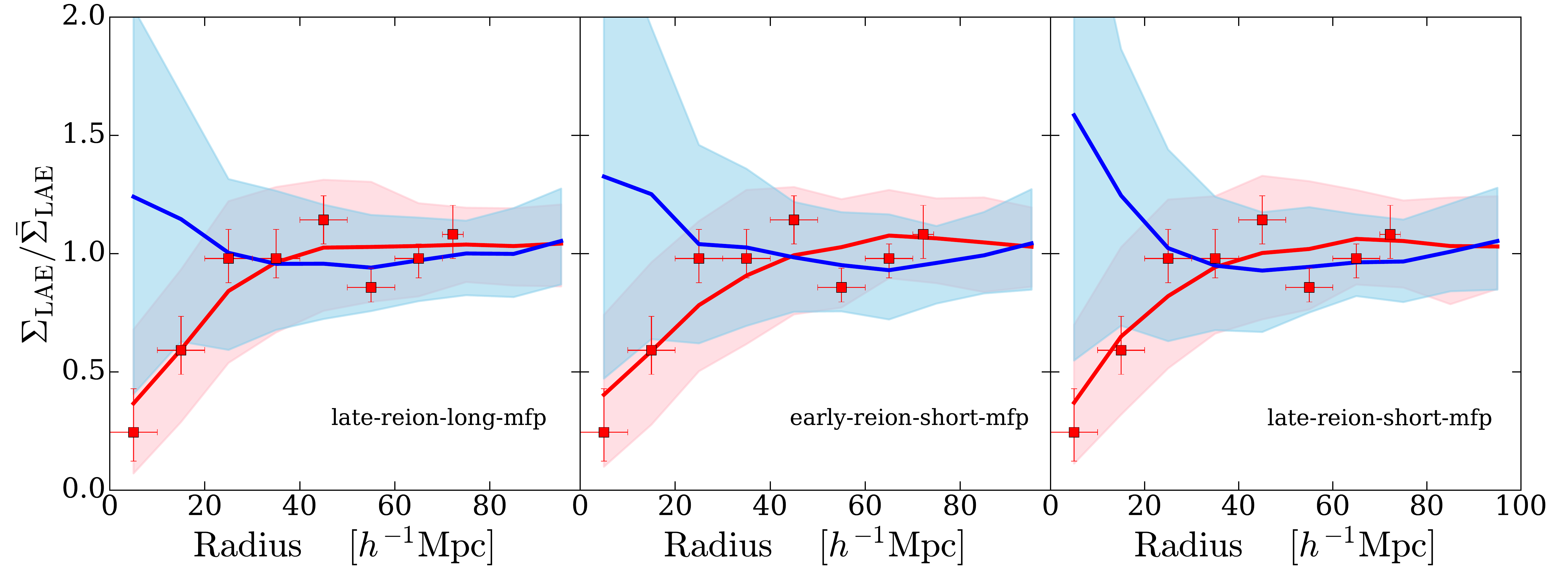}}
\hspace{-0.28cm}
\end{center}
\caption{LAE surface densities as a function of radius from transmissive (blue) and opaque (red) sight lines. Surface densities are reported in units of the mean. We select 100 sight lines with the lowest \taueff\ (blue) and 100 with the longest Ly$\alpha$ troughs (red).  The curves show the mean surface densities and the shadings show the 10th and 90th percentiles.  For reference, the data points show the measurements of \citet{Becker_2018ApJ} for the extreme Ly$\alpha$ trough in the sightline of J0148.      }
\label{FIG:LAE_surface_densities}
\end{figure*}

\begin{figure}
\begin{center}
\resizebox{8.5cm}{!}{\includegraphics{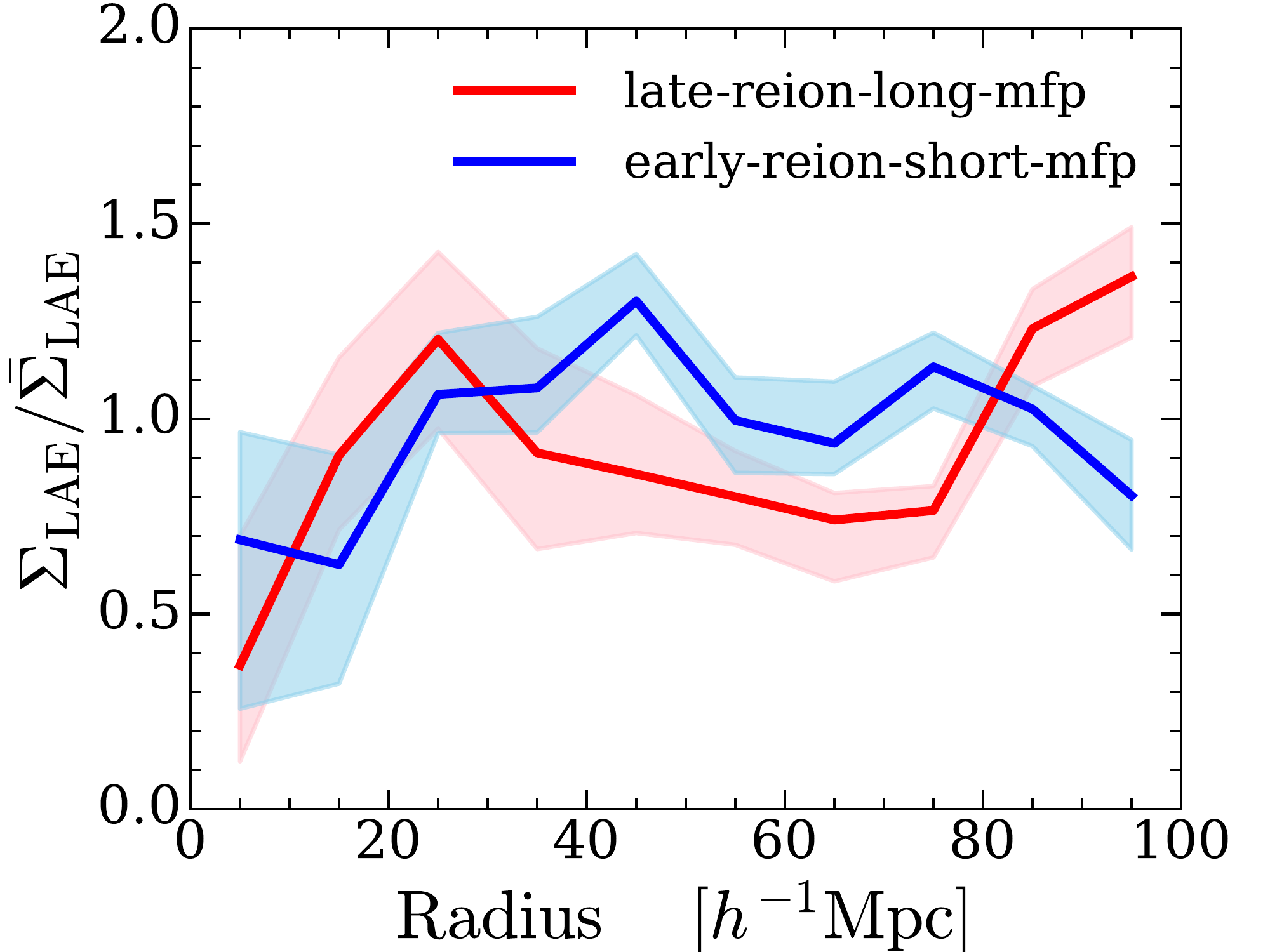}}
\end{center}
\caption{Example sight lines with low \taueff, but displaying an under-density of LAEs.  The shadings span 10 realizations of the LAE population.   For both models, we find $\approx12$ out of 100 sight lines with $\Sigma_{\rm LAE}/\bar{\Sigma}_{\rm LAE} < 0.5$ at $R=5$\mpc, indicating that these sight lines are relatively rare. Sight lines of this nature intersect warm voids that were more recently reionized.      }
\label{FIG:transmissive_deficit}
\end{figure}

\begin{figure}
\begin{center}
\resizebox{8.5cm}{!}{\includegraphics{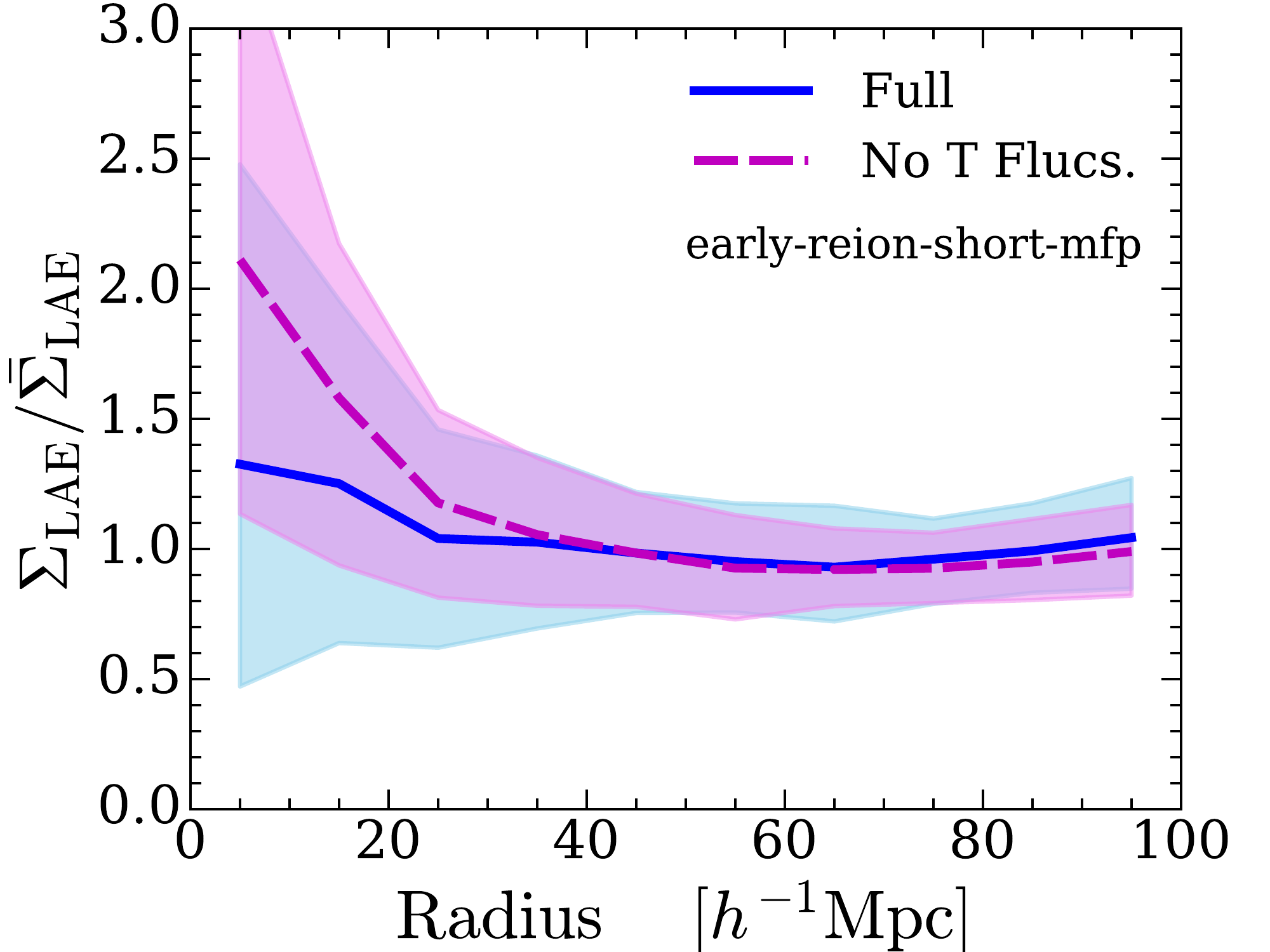}}
\end{center}
\caption{ Effect of temperature fluctuations from reionization on the correlation between low-\taueff and LAE density.  For the magenta/dashed curve, we remove the reionization temperature fluctuations and use the original temperatures from our hydro simulation.  This effectively assumes that reionization ended early enough for the temperature-density relation to relax to a tight power-law form.  Relic temperature fluctuations from reionization create more low-\taueff\ regions out of voids.    }
\label{FIG:early-short_noTflucs}
\end{figure}

Here we mock up LAE surveys along opaque and transmissive forest sight lines in the vein of the recent Hyper Suprime-Cam (HSC) survey towards J0148 by \citet{Becker_2018ApJ}.  For each of our three models, we build a sample of the most opaque and transmissive sight lines by identifying the 100 longest \lya troughs, and the 100 most transmissive (lowest \taueff) segments of length $50$\mpc.  For simplicity, in this section we neglect time evolution along the light cone and extract sight lines from our $z=5.8$ snapshot.  This simplification is unlikely to alter our statistical conclusions on the correlation with LAE surface density.  

We adopt a procedure similar to that in \citet{Keating_2019arXiv} to generate mock LAE surveys along the extracted sight lines.   We abundance match the dark matter halos in our simulation using the UV luminosity function of \citet{Bouwens_2015ApJ_lim}.  We then employ the empirically calibrated model distribution of \citet{2012MNRAS.419.3181D} to draw rest-frame equivalent widths (REWs) for each halo.  The intrinsic spectrum of each galaxy is modeled as a power-law continuum of the form $F_\lambda\ \propto \lambda^{-2}$, plus a double-peaked \lya emission line. Following \citet{2018MNRAS.479.2564W}, we assume that the component of the emission line blue-ward of systemic is fully attenuated, and we model the red-ward component as a Gaussian with offset of $100~\mathrm{km/s}$ and width $\sigma = 80~\mathrm{km/s}$.  We trace skewers of length $150$\mpc from each halo and attenuate the mock spectra according to the opacities along the sight lines.  For the late-reionization models, this includes accounting for the damping wing attenuation from neutral hydrogen along the sight line.  We then measure the HSC $NB816$ and $i2$ magnitudes using the published transmission curves.  We apply two of the three cuts used by \citet{Becker_2018ApJ}, $NB816 \leq 26.0$ and $i2 - NB816 \geq1.2$, neglecting as in \citet{Keating_2019arXiv} the $r2$ magnitude cut which extends to redshifts below our modeling.  Lastly, as the forest sight lines are drawn at random angles through our box, we rotate the mock LAEs to a head-on view and then bin by transverse radius to measure the surface density profile. This procedure yields a mean LAE surface density of $\approx 0.05~(h^{-1}\mathrm{Mpc})^{-2}$, in good agreement with the observed mean surface density in \citet{Becker_2018ApJ}.  

In Fig. \ref{FIG:LAE_surface_densities} we show the mean LAE surface density profiles (averaged over 100 sight lines).  The red and blue curves correspond to extreme troughs and transmissive segments, respectively.  The shading shows the 10th and 90th percentiles in our ensembles.  For reference, the red data points show the \citet{Becker_2018ApJ} measurements for the J0148 trough.  For all three models, the surface density profiles towards troughs are consistent with the measurements \citet{Becker_2018ApJ}, reflecting the fact that extreme troughs are generated by voids in these models.   In all three models the transmissive regions display a larger scatter in surface densities. More often than not, however, the most transmissive segments of the forest correspond to over-densities in LAEs, even in the {\UrlFont late-reion-long-mfp} model.  One might have guessed that the most transmissive regions in the {\UrlFont late-reion-long-mfp} model are typically hot, recently reionized voids.  Our results suggest that the situation is more complicated when $\Gamma_{\rm HI}$ and temperature fluctuations, and neutral islands, are all at play.  

\citet{Keating_2019arXiv} suggested that the most transmissive sight lines may offer a path to discriminating between models.  Whereas the most transmissive (low \taueff) sight lines typically correspond to over-densities in the model of \citet{Davies_2016MNRAS}, in a late-reionization scenario, at least some of the most transmissive regions should correspond to hot, recently reionized voids.  However, \citet{Keating_2019arXiv} did not quantify how common such transmissive voids are.  In our  {\UrlFont late-reion-long-mfp} model we do find transmissive sight lines with a deficit of LAEs, much like the one reported in \citet{Keating_2019arXiv}, but they are relatively rare.  For example, 12 out of the 100 transmissive sight lines in our sample have surface densities $\Sigma_{\rm LAE} / \bar{\Sigma}_{\rm LAE} < 0.5$ at $R=5$ \mpc.  One such sight line is shown as the red curve in Fig. \ref{FIG:transmissive_deficit}, where the shading spans ten realizations of the LAE population (i.e. ten different samples drawn randomly from the REW distribution).  However, we also find these under-dense sight lines with similar frequency (13/100) in the {\UrlFont early-reion-short-mfp} model -- one of which is shown as the blue curve in Fig. \ref{FIG:transmissive_deficit}. 

At face value, the wide dispersion for transmissive sight lines in our {\UrlFont early-reion-short-mfp} model appears inconsistent with the tighter correlation between transmission and LAE over-density reported by \citet{Davies_2018ApJ}.  We find that this discrepancy owes to the presence of temperature fluctuations in our {\UrlFont early-reion-short-mfp} model.  To demonstrate this, we remove the reionization temperature fluctuations from the model and use the original temperatures from the hydro simulation.  Again, this effectively assumes that reionization was completed early enough for the temperature-density relation to relax to a tight power-law form. The results of this exercise are shown as the magenta/dashed curve in Fig. \ref{FIG:early-short_noTflucs}.  When we remove the temperature fluctuations we find a tighter correlation between transmission and LAE density, more consistent with the findings of \citet{Davies_2018ApJ}, who did not include temperature fluctuations from reionization in their model.

Our results suggest that it is difficult to discern our three models by targeting the most transmissive segments of the \lya forest.  This difficulty arises from the interplay between $\Gamma_{\rm HI}$ and temperature fluctuations in generating large scatter among the low-\taueff\ sight lines in all three models.  
On the other hand, if a strong correlation between low-\taueff\ and galaxy/LAE over-densities is measured (ruling out all three of these models), it may be indicative of a relaxed temperature-density relation with minimal scatter, and therefore an earlier end to reionization.  Surveys targeting \lya forest sight lines thus remain a potentially insightful window into the physical nature of the fluctuations and ultimately of reionization.  


\section{Conclusions}

\label{sec:conclusions}

Previous studies have noted a marked rise in the scatter of Ly$\alpha$ forest opacities at $z>5.5$.   We explored a late-reionization model, in which the neutral fraction is still $\approx 10\%$ at $z=5.5$, and found that it can naturally account for the scatter, including long Ly$\alpha$ troughs with properties similar to the one observed in J0148 by \citet{Becker_2015MNRAS}.  We attempted to contrast this model with two competing models: (1) An ``early" reionization scenario characterized by large fluctuations in the post-reionization UVB.  This model is similar to that of \citet{Davies_2016MNRAS}; (2) A separate scenario that blends elements of the \citet{Davies_2016MNRAS} model with late reionization.  The nature of Ly$\alpha$ troughs is different between the models.  In our fiducial late-reionization model, every long trough contains neutral islands because they are necessary for generating such large-scale regions with high opacity.   However, in the early reionzation scenario, long troughs can occur with or without neutral islands.  The opacity fluctuations are instead driven by post-reionization spatial variations in the mean free path and UVB.

We compared Ly$\alpha$ and Ly$\beta$ forest statistics such as the distributions and correlations of \taueff, trough size distributions, and dark pixel fractions, to look for easy discriminators between the models.  We found that the models are broadly similar in these statistics, with the exception of the frequency of intermediate-size ($5 \lesssim L \lesssim 20$ \mpc) troughs in the Ly$\beta$ forest.  Our early reionization model yields fewer gaps in the Ly$\beta$ forest of these sizes.  Motivated by the recent findings of \citet{Becker_2018ApJ}, we also explored LAE surveys centered on the most opaque, and most transmissive, regions of the $z = 5.7$ forest.  All of our models are consistent with the measured LAE surface densities around the Ly$\alpha$ trough of J0148.  The predicted surface densities towards the most transmissive sight lines exhibit a large scatter owing to the interplay between UVB and temperature fluctuations. The amount of scatter is similar between the three models, rendering it difficult to distinguish them with this method.  However, if observations instead reveal that transmissive sight lines are overwhelmingly associated with overdensities (ruling out all three of our models), this may indicate a more subdued role for temperature fluctuations, e.g. due to an earlier reionization.  The correlation between forest transmission and galaxies/LAEs is an informative window into the nature of IGM fluctuations at these redshifts \citep{Becker_2018ApJ, 2018MNRAS.479...43K, 2019arXiv190909077K}.       

Among the tests that we considered, perhaps the most promising observational window is the thermal history of the IGM.  Models in which reionization ended at $z=5-5.5$ predict significantly hotter temperatures at those redshifts compared with earlier reionization models.  On the other hand, the gas density fields exhibit a higher degree of pressure smoothing in the latter.  In principle these differences should be detectable in the small-scale structure of the forest if measurements can be extended to $z\approx 5.5$ with high-resolution ($\Delta v \lesssim 30$ km/s), high signal-to-noise ($S/N > 15$) quasar spectra.      One caveat here is the uncertainty in the contribution of quasars to the UVB at these redshifts, which we have not modeled here.  A larger-than-expected quasar contribution can boost temperatures by \HeII\ ionizations, complicating the interpretation for the timing of hydrogren reionization \citep{D'Aloisio_2017MNRAS}.  

Another path to probing the nature of the UVB fluctuations is to expand measurements of the mean free path at $z>5$.  Reproducing the extreme forest opacities in a model along the lines of \citet{Davies_2016MNRAS} requires that the global mean free path be at least a factor of 2 lower than current measurements at $z< 5.2$ suggest.  Such a discrepancy is possible -- albeit unlikely -- if the existing mean free path measurement at $z\approx 5.2$ is biased significantly by the quasar proximity effect \citep{D'Aloisio_2018MNRAS}.  Quasars of lower luminosity would be less impacted by the effect, providing in principle a straightforward way to test for the bias.  Because one of the primary differences between our models is the mean free path, we argue that extending direct measurements to include fainter quasars would be another informative step towards diagnosing the origin of the forest opacity fluctuations.  

Lastly, our results motivate future improvements to the modeling.  The largest differences between earlier and later reionization models probably lie in the small-scale structure of the forest, which we have explored only indirectly here.  Future progress requires simulations with larger dynamic range, and that capture the dynamical response of the gas.  The current work represents an early step towards exploring the testability of these models.


\appendix
\section*{Acknowledgments}

The authors thank George Becker, Matt McQuinn, Fred Davies, Hy Trac, Andrei Mesinger, and the anonymous referee for helpful discussions and comments on this manuscript. The authors are especially grateful to Hy Trac for providing the RadHydro code, and to Elisa Boera for providing the posterior distributions in Fig. \ref{fig:thermal_hist_cont}.  The authors acknowledge support from HST award HST-AR15013.005-A.  Computations were made possible by NSF XSEDE allocation TG-AST120066.

\bibliographystyle{mnras}
\bibliography{bibliography}

\end{document}